%% file: redmapper_centering.tex
\documentclass[fleqn,usenatbib]{mnras}
\usepackage{graphicx}
\usepackage{amsmath}
\usepackage{color}
\usepackage{comment}
\usepackage{rotating}
\usepackage{lineno}
\usepackage{ae,aecompl}
\usepackage{mathptmx}
\usepackage{txfonts}
\usepackage[T1]{fontenc}

\newcommand{\hinv}{\ensuremath{\, h^{-1}}}%
\newcommand{\msol}{\ensuremath{\, {\rm M}_\odot}}         
\newcommand{\kpc}{\ensuremath{\, {\rm kpc}}}         
\newcommand{\mpc}{\ensuremath{\, {\rm Mpc}}}         
\newcommand{\gpc}{\ensuremath{\, {\rm Gpc}}}%

\newcommand{\rtwoh}{R_{\rm 200c}}

\newcommand{\red}[1]{\textcolor{red}{#1}}

\usepackage{eso-pic}

\AddToShipoutPictureBG*{%
  \AtPageUpperLeft{%
    \hspace{0.75\paperwidth}%
    \raisebox{-3.5\baselineskip}{%
      \makebox[0pt][l]{\textnormal{DES 2017-0316}}
}}}%

\AddToShipoutPictureBG*{%
  \AtPageUpperLeft{%
    \hspace{0.75\paperwidth}%
    \raisebox{-4.5\baselineskip}{%
      \makebox[0pt][l]{\textnormal{FERMILAB-PUB-19-013-AE-PPD}}
}}}%

\newcounter{RMDone}
\def\RM{\ifthenelse{\equal{\arabic{RMDone}}{0}}{redMaPPer \setcounter{RMDone}{1}}{redMaPPer}}

\title[RedMaPPer Mis-centering]{Dark Energy Survey Year 1 Results: Calibration of Cluster Mis-centering in the redMaPPer Catalogs}
\input{centering_authors.tex}

\date{Accepted XXX. Received YYY; in original form ZZZ}

\pubyear{2019}

\begin{document}

\label{firstpage}
\pagerange{\pageref{firstpage}--\pageref{lastpage}}

\maketitle
\begin{abstract}

The center determination of a galaxy cluster from an optical cluster finding algorithm can be offset from  theoretical prescriptions or $N$-body definitions of its host halo center.  These offsets impact the recovered cluster statistics, affecting both richness measurements and the weak lensing shear profile around the clusters.  This paper models the centering performance of the \RM~cluster finding algorithm using archival X-ray observations of \RM-selected clusters. Assuming the X-ray emission peaks as the fiducial halo centers, and through analyzing their offsets to the \RM~centers, we find that $\sim 75\pm 8 \%$  of the \RM~clusters are well centered and the mis-centered offset follows a Gamma distribution in normalized, projected distance. These mis-centering offsets cause a systematic underestimation of cluster richness relative to the well-centered clusters, for which we propose a descriptive model. Our results enable the DES Y1 cluster cosmology analysis by characterizing the necessary corrections to  both the weak lensing and richness abundance functions of the DES Y1 redMaPPer cluster catalog.

\end{abstract}

\section{Introduction} \label{sec:intro}

The abundance of galaxy clusters is a sensitive probe of cosmological models \citep[see reviews and the referenced literature in][]{2011ARA&A..49..409A, 2013PhR...530...87W}. Cluster cosmology studies from the latest optical imaging surveys such as the Dark Energy Survey \citep{DES2018} will deliver significant improvement in precision over previous studies based on the Sloan Digital Sky Survey \citep[SDSS,][]{2010ApJ...708..645R} and require \citep{McClintock2018} accurate understanding of various systematic effects such as the orientation of clusters \citep{2012MNRAS.426.1829N, 2014MNRAS.443.1713D}, correlated structures and their projection effect \citep{Erickson2011, 2018arXiv180707072C}, mass profile modeling uncertainties \citep{McClintock2018}, and the contamination of cluster member galaxies in the lensing measurements \citep{Varga:2018}.

One such important systematic effect is the mis-identification of cluster centers 
\citep[][]{2007arXiv0709.1159J, 2007ApJ...656...27J, 2017MNRAS.469.4899M, 2017MNRAS.466.3103S, McClintock2018, DES2018}.
Cluster observables, e.g.~gravitational shear profiles, must be compared to models in order to derive constraints on parameters, and the models are based on some definition of cluster center, described theoretically or on the basis of the matter density field in $N$-body simulations.  Cluster lensing studies based on data from DES \citep{McClintock2018} require accurate knowledge of cluster mis-centering fraction and offset distribution in order to forward model masses using analytic halo profiles.

Optical cluster finders often attempt to identify a central galaxy as the center \citep{2007ApJ...660..239K, 2010ApJS..191..254H, 2014ApJ...785..104R, 2017arXiv170100818O}. These central galaxies are typically quenched of star formation activities and may appear to be the brightest galaxy in a cluster \citep[see studies in][]{2011MNRAS.410..417S, 2015MNRAS.452..998H, 2014ApJ...797...82L}. The identification of cluster central galaxies may seem straightforward given their dominant appearances, but mis-identifications or offsets relative to any other theoretical definition of center is inevitable.  
Because massive halos experience growth through mergers, cluster central galaxies can be displaced from the local gravitational potential minimum \citep[e.g., ][]{Martel2014}.  A related effect is when a second galaxy within the halo is chosen as the center by the cluster finding algorithm.  
For a color-based scheme focused on the reddest galaxies, this may happen if the central galaxy of the host halo has experienced recent star formation \citep[e.g., ][]{2012Natur.488..349M, Donahue2015} or if a merging event brings in two nearly identical central galaxies of the progenitor halos, as in the case of the Coma cluster \citep[e.g., ][]{2001ApJ...555L..87V}.  
Another cause of mis-centering is when galaxies lying outside the primary host halo, but aligned in projection, are chosen as the central galaxy by the cluster finding algorithm.

The centering performance of optical cluster finding algorithms has been characterized with various methods. 
Cluster hot gas is an excellent tracer of the cluster potential as the dominant baryonic mass component. Cluster X-ray or thermal Sunyaev-Zel'dovich (tSZ) observation centers, identified as the centroids or the peaks of the surface brightness, are often used to calibrate the optically selected centers (see examples of X-ray studies in \citealt{2004ApJ...617..879L, 2012MNRAS.422.2213S, 2013ApJ...767..116M, 2014ApJ...783...80R, 2014ApJ...797...82L, 2016ApJS..224....1R, 2016ApJ...816...98Z} and examples of tSZ studies in \citealt{2012ApJ...761...22S, 2015MNRAS.454.2305S}). Other than calibration to centers identified in multi-wavelength observations, cluster centering has been characterized through comparing cluster radial profiles (lensing or galaxy number count) to those of a cluster sample with well-known centers from X-ray or optical data \citep{2018MNRAS.480.2689H, 2017arXiv171209030L} and through examining the velocity and separation distribution of cluster satellite galaxies \citep{2011MNRAS.410..417S}.

In this paper, we characterize the centering performance of the \RM~cluster finding algorithm -- a method for identifying galaxy clusters from optical imaging data. The \RM~algorithm excels in producing a complete and efficient cluster sample with an accurate richness mass proxy and precise redshift estimations, as characterized with multi-wavelength and spectroscopic data \citep{2012ApJ...746..178R, 2014ApJ...783...80R, 2015MNRAS.450..592R, 2015MNRAS.453...38R, 2015MNRAS.454.2305S, 2017arXiv170701907M}. Cluster catalogs constructed from SDSS \citep{2014ApJ...785..104R} and DES data \citep{2016ApJS..224....1R} are used to derive cosmological constraints in \cite{Costanzi2018, DES2018}. In terms of the cluster center identification, \RM~is not exempt from occasional mis-identification of the central galaxy.

The centering distribution of the \RM~algorithm has been studied using almost all of the aforementioned methods \citep{2014ApJ...783...80R, 2015MNRAS.454.2305S, 2016ApJS..224....1R, 2018MNRAS.480.2689H}. In this paper, we model the cluster centering distribution with 211 high signal-to-noise X-ray cluster detections associated with the \RM~SDSS DR8 and DES Year 1 samples from the Chandra public archives, and the model constraints are then validated with X-ray cluster detections from the XMM public archives. We focus on the modeling aspects of \RM~centering performance in this paper, while the X-ray data processing procedures are presented in \cite{Hollowood2018} and Giles et al. in prep. The data set and methods we employ allow us to analyze the \RM~centering performance with the highest precision to date. 
We also develop a model to characterize how mis-centering affects \RM~richness estimation and discuss the impact of cluster mis-centering on cluster cosmology analyses. It is the first time that this effect has been quantified. 

This paper is a companion paper to the DES and SDSS cluster weak lensing and cosmology studies presented in \cite{McClintock2018}, \cite{Costanzi2018} and \cite{DES2018}. It uses similar data products to \cite{Farahi2018}.

Throughout this paper, we assume a Flat $\Lambda$CDM cosmology with $h =$ 0.7 and $\Omega_m=0.3$. 

\section{Data}\label{sec:data}

\subsection{The redMaPPer Catalogs}\label{sec:redmapper}

The redMapper algorithm examines galaxy color, spatial over-density and galaxy luminosity distribution to identify possible galaxy clusters. Cluster centers are placed on a central galaxy candidate according to the color, luminosity and galaxy over-density computed around the galaxy. Up to five central galaxy candidates are recorded for each cluster with probabilities assigned to them. The cluster center is chosen to be the most probable one. The \RM~algorithm also estimates a richness as a mass proxy, $\lambda$, which is a probabilistic count of red sequence galaxies within an aperture centered on the central galaxy candidate. Detailed presentation of this algorithm can be found in \cite{2014ApJ...785..104R, 2016ApJS..224....1R}. 

The \RM~samples studied in this paper have been derived from both SDSS \citep{2014ApJ...785..104R, 2016ApJS..224....1R} and DES \citep{2016ApJS..224....1R, McClintock2018} data. We use the \RM~SDSS 6.3.1 sample \citep{Costanzi2018}   derived from SDSS DR8 \citep{2011ApJS..193...29A} photometric data and the \RM~DES-Y1-6.4.17 volume-limited catalog which is based on the DES Y1 gold catalog \citep{2017arXiv170801531D}, with cluster richnesses $\ge$ 20. For the SDSS \RM~sample, we consider clusters in the redshift range of 0.1 to 0.35 which are nearly volume-limited. For the DES \RM~sample, we select clusters between redshifts 0.2 and 0.7. 

\subsection{Comparison of Centers in redMaPPer Catalogs}

In this paper, we treat the DES and SDSS \RM~catalogs as two independent samples and characterize their centering performances separately. We examine the offset distribution between redMaPPer centers of the overlapping clusters in the DES and SDSS samples, to estimate an upper limit of the well-centered cluster fraction.  

We match between the DES and SDSS \RM~samples to identify the overlapping clusters in the redshift range of 0.2 to 0.35. For a pair of DES and SDSS clusters to be considered a match, their redshift difference, $|\Delta z|$, must be less than 0.05 to account for the scatter in photometric redshifts and possible blending effects. The \RM~centers must be within 1 $R_\lambda$ ($R_\lambda =(\lambda/100)^{0.2}  h^{-1}\mpc$). The radius aperture is  chosen because \RM~does not consider the clusters to be the same if their centering offset is larger than $R_\lambda$. The richness estimations derived from SDSS and DES data have an average relation of $\lambda_{DES} = (0.88\pm0.03)\times\lambda_{SDSS}+(3.28\pm1.20)$. We look for matches to $\lambda > 20$ DES clusters in the SDSS~\RM~sample by lowering the SDSS~\RM~$\lambda$ threshold to $ 5$ to account for the $\lambda$ difference and scatter between DES and SDSS.

There are 150 DES clusters with $\lambda>20$ and 38,786 SDSS clusters with $\lambda>5$ in the overlap region of both catalogs after applying redshift, position, and mask cuts. Of these 150 DES clusters, 148\footnote{After further investigation, the other 2 DES clusters appear to have SDSS matches with the same central galaxy selections, but the redshift differences between the DES and SDSS match are $\sim$ 0.06, and thus didn't pass our strict redshift difference cuts.} have SDSS matches given the criteria listed above. 15 of these 148 clusters have at least two matches and the most likely SDSS match was chosen by inspection of redshift, position, and richness. As the purpose of this matching process is to verify the centering consistency in SDSS/DES, we further remove 3 clusters in the total sample (148) because of our poor confidence in the match: the SDSS matching candidates have large richness differences with their respective DES clusters. This further reduces our matching sample size to 145. 

\begin{figure}
\includegraphics[width=0.95\linewidth]{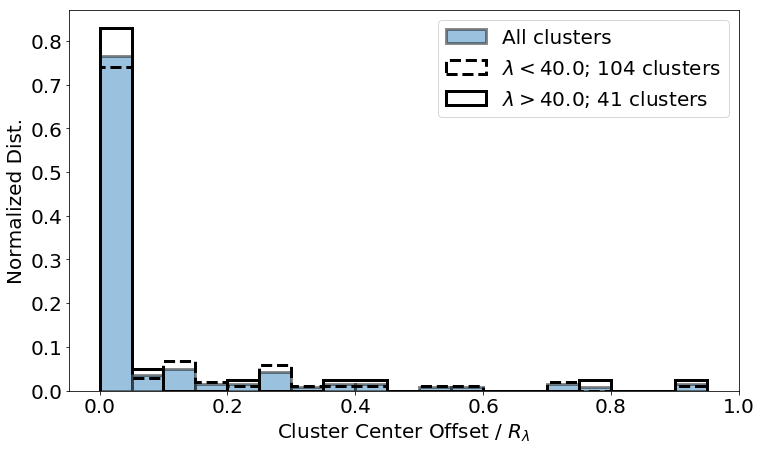}
\caption{The offset distribution of cluster centers assigned in the SDSS and DES redMaPPer catalogs, measured in radial units of $R_\lambda$ ($R_\lambda =(\lambda/100)^{0.2}  h^{-1}\mpc$, which is $\sim$ 1 Mpc for $\lambda$ of 20). The solid blue histogram is the full sample of overlap clusters while the solid and dashed histograms are the same distribution in richness bins of $\lambda>40$ and $\lambda<40$ respectively. 77\% of clusters have offsets below 0.05 $R_\lambda$ and are considered self-consistent, with the remaining 23\% comprising a long tail. We notice a marginal richness dependence that richer clusters appear to be more consistently centered in DES and SDSS. }
\label{fig:offset_compare_dist}
\end{figure}

Figure \ref{fig:offset_compare_dist} shows the scaled offset distribution between SDSS and DES centers for the matched clusters. For 77\% of the matched clusters, their SDSS and DES centers are within 0.05 $R_\lambda$ \citep[corresponding to $\sim$ 50 $\mathrm{kpc}$ at $\lambda=20$, close to the size of a typical central galaxy, ][]{2005MNRAS.358..949Z,2011MNRAS.414..445S}. We consider these clusters as consistently centered between the two catalogs. The remaining 23\% of the clusters comprise a long tail in the SDSS and DES offset distribution up to 1 $R_\lambda$. The inconsistency indicates that for at least one of the samples, the mis-centered fraction is $\ge 0.23/2=0.115$. As we do not have sufficient information (i.e., enough X-ray observations) in the DES/SDSS samples to analyze which has a greater rate of well-centering,  we decide to {\it independently analyze the SDSS and DES \RM~samples in this paper}. 

We further examine the richness distribution of the offsets by dividing the clusters into two richness ranges. We do not notice a significant richness trend -- 34 out of 41 clusters at richness above 40 are consistently centered {\it VS} 77 out of 101 at richness below 40.

\subsection{Chandra X-ray Data and Center Measurement}
\label{sec:xray}
\label{sec:chandra-redmapper}


While the comparison of centering between different redMaPPer catalogs above gives an indication of the minimum level of mis-centering, we use X-ray data to calibrate the absolute value of the well-centered cluster fraction, and the offset distribution of mis-centering clusters in each of the redMaPPer samples. 

In this paper, we use cluster X-ray emission peaks as the fiducial centers and rely on these to estimate the \RM~mis-centering fraction and offsets.  
Some previous studies have used the X-ray and tSZ centroids within different aperture sizes that closely resemble the centroids of the cluster gravitational potential to calibrate the cluster centering distribution \citep{2012MNRAS.422.2213S, 2012ApJ...761...22S, 2015MNRAS.454.2305S, 2016ApJ...816...98Z}, while others use X-ray emission peaks that closely resemble the peaks of the cluster matter distribution \citep{2004ApJ...617..879L, 2013ApJ...767..116M, 2014ApJ...797...82L}. 
In DES and SDSS cluster weak lensing and cosmology studies \citep{McClintock2018, Costanzi2018, DES2018}, the aim is to quantify redMaPPer's accuracy in identifying the galaxy near the center of a cluster's host dark matter halo, or the galaxy that corresponds to the density peak of the dark matter halo \citep{2008ApJ...688..709T}. 
To this end, we employ the X-ray peak position as a proxy for the host halo center, and measure the distribution of the projected offsets between X-ray peaks and redMaPPer central galaxies. 

We search for X-ray observations and determine X-ray peaks in archival Chandra data for redMaPPer clusters in both SDSS and DES Y1 of richness above 20. RedMaPPer clusters falling within an archival Chandra observation are analyzed with a custom pipeline MATCha, described in \citet{Hollowood2018}.  A summary of the X-ray analysis follows.

\begin{figure}
 \centering
 \includegraphics[width=0.95\linewidth]{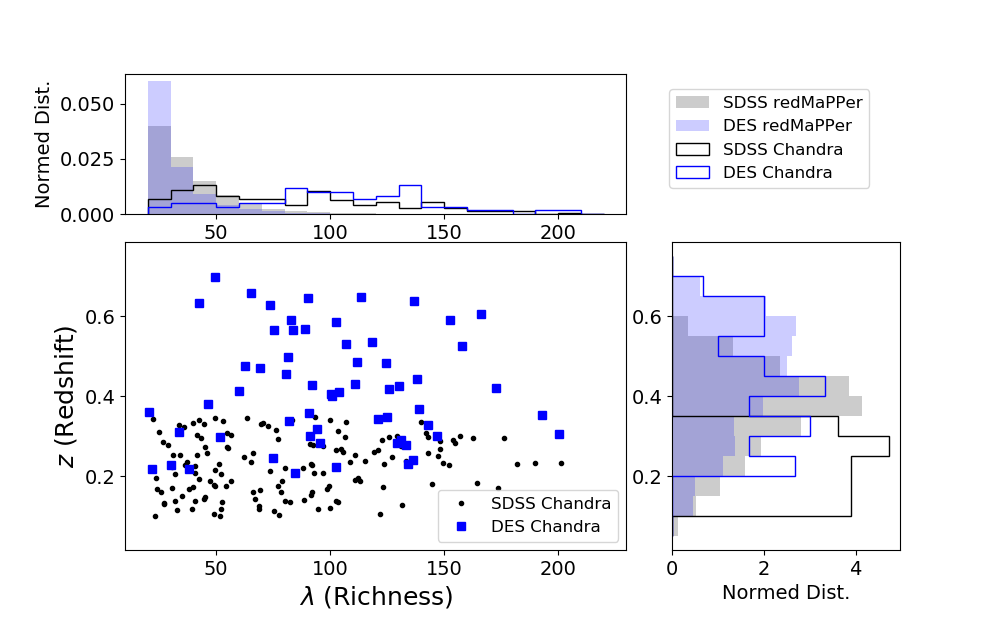}
 \caption{Normalized redshift and richness distributions of the SDSS (black) and DES (blue) redMaPPer clusters matched to archival Chandra observations. The X-ray matched clusters have much higher richnesses than the general redMapper sample, although we do not find significant richness dependence of the results presented in the paper. } 
 \label{fig:xray-redmapper_c}
\end{figure}

\begin{figure*}
\includegraphics[width=0.95\linewidth]{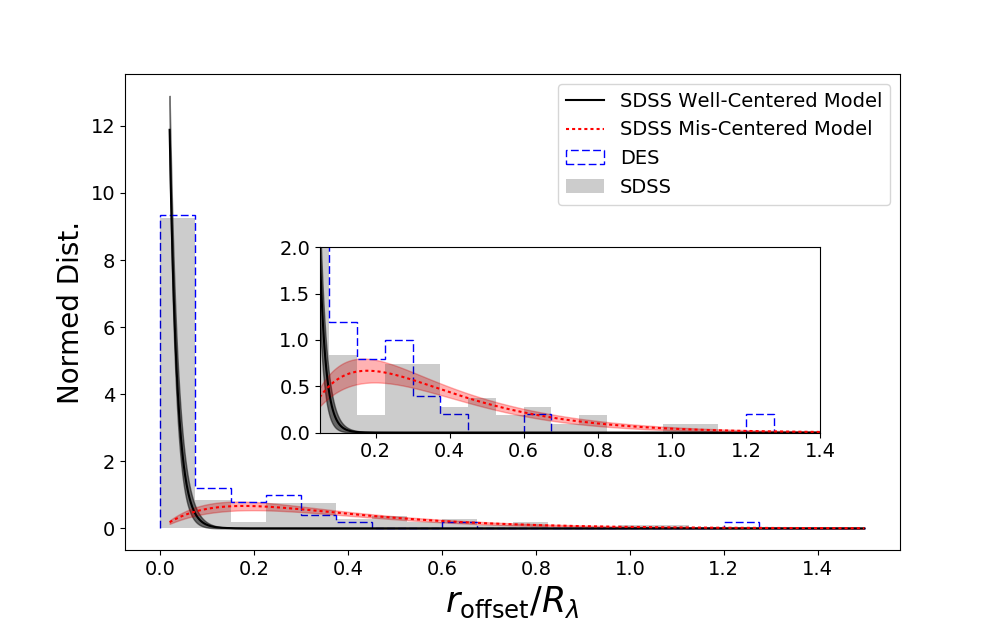}
\caption{The $R_\lambda$ ($R_\lambda =(\lambda/100)^{0.2}  h^{-1}\mpc$) scaled offset distribution between the redMaPPer centers and the X-ray emission peaks for the redMaPPer SDSS and DES samples from the Chandra archival observations, with the inset zooming on the mis-centered component, starting at $r_\mathrm{offset}/R_\lambda=0.05$. The distribution can be fitted with two components -- a concentrated component that represents the well centered redMaPPer clusters, and an extended component that represents the mis-centered redMaPPer clusters. The best fit SDSS offset model is shown as the solid lines (black: well-centered model, red: mis-centered model), with the shaded regions representing the uncertainties. }
\label{fig:xrayoffset1}
\end{figure*}

For each of the \RM~clusters with archival Chandra observations, starting with an initial aperture of 500 kpc radius centered on the \RM~center, the pipeline first determines the X-ray centroid, re-centers, and then iteratively finds X-ray centroids until convergence is reached within 15 kpc. A cluster is considered to be detected if the signal-to-noise ratio within a final 500 kpc aperture centered on the converged centroid is greater than 5.  For detected clusters, MATCha analysis proceeds with attempts to measure $L_X$, $T_X$, and centroid within a set of apertures including 500 kpc, $r_{2500}$, $r_{500}$, and core-cropped $r_{500}$.
Visual checks of non-detected \RM~clusters are employed to examine if any \RM~clusters with Chandra observations were omitted in the process. We find one SDSS \RM~cluster with a large offset of 1.82 Mpc (1.40 $R_\lambda$ for this cluster) between the X-ray centroid and the \RM~center, possibly over-looked because of the initial 500 kpc X-ray centroid searching criteria. Since this omission makes up less than 1\% of the total SDSS Chandra sample, we do not consider it in further analyses. No similar cases were found in the DES Y1 sample.

For the clusters with X-ray detections, MATCha additionally determines the position of the X-ray peak starting from the reduced, exposure-corrected, and point source subtracted images.  Images are smoothed with a Gaussian with $\sigma = 50$ kpc width, and the peak is defined to be the brightest pixel in this smoothed image.  All peaks are then visually examined.  In a small number of cases relic point source emission or the removal of a point source near the cluster peak are found to bias the peak determination.  The peak position is adjusted after accounting for the point source emission.  In addition, two failure modes are flagged and removed from the sample.  First, for the centering analysis we remove clusters falling on or near a chip edge in the X-ray observation such that the position of the X-ray peak could not be reliably determined.  Second, in a few cases the identified X-ray cluster is clearly not the redMaPPer cluster (e.g. a bright foreground or background cluster in the same observation), and these clusters are likewise removed (see \citealt{Hollowood2018} for further detail).
 Moreover, there are some special \RM~mis\-centering cases, denoted as mis-percolations in \cite{Hollowood2018}, because these cases are related to a "percolation" procedure of redMaPPer. In these cases there is a spatially close pair of clusters with similar redshifts, and the one with a less 
luminous X-ray detection is assigned with a greater richness. \cite{Hollowood2018} manually associates the richer \RM~candidate with the more luminous X-ray detection and removes the less rich system from the X-ray samples.

Finally, to improve the accuracy of X-ray peak location, among all the clusters identified in \citealt{Hollowood2018}, we further impose a signal-to-noise cut removing clusters with a signal-to-noise ratio less than 6.5 within a 500 kpc aperture. In the end, 144 \RM~SDSS clusters are identified with X-ray peak centers in the Chandra archival data.

The compilation of the DES \RM~Chandra sample follows a similar process to the SDSS \RM~sample, with the exception that the X-ray peaks of the DES sample are initially identified around the \RM~centers within 500 kpc due to pipeline re-factoring. The peak identifications are visually examined and adjusted if needed. Overall, 67 DES \RM~clusters are identified with Chandra observations of signal-to-noise ratio higher than 6.5. 

In Figure~\ref{fig:xray-redmapper_c}, we show the redshift and richness distributions of the SDSS and DES \RM~Chandra samples. Figure~\ref{fig:xrayoffset1} shows the scaled offset distributions between the X-ray peaks and the \RM~centers. The tail of the offset distribution indicates a population of mis-centered clusters. Examples of the mis-centered clusters can be found in \citet{Hollowood2018}.

\section{The X-ray redMapper Offset}
\label{sec:center}

\subsection{Model}
\label{sec:centermodel}

\begin{figure}
\includegraphics[width=0.95\linewidth]{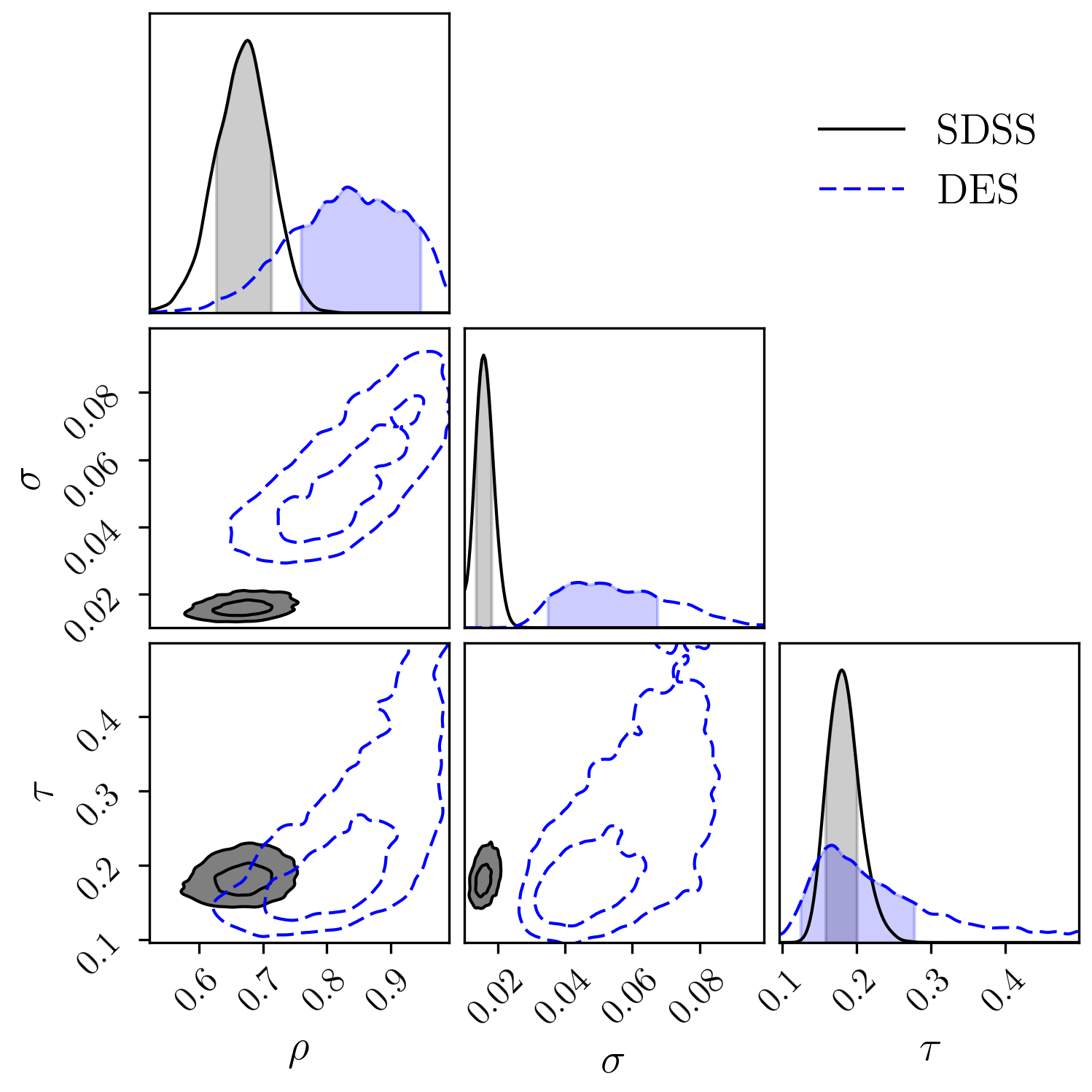}
\caption{Centering offset parameter constraints (Equation~\ref{eq:offset}) for the Chandra DES (blue) and SDSS (gray) redMaPPer samples. About $70\%$ of the DES and SDSS redMaPPer clusters appear to be well centered in both samples (indicated by the $\rho$ parameter). For the mis-centered clusters, their mis-centering offsets is characterized by a Gamma distribution with a characteristic offset (the $\tau$ parameter) around 0.18 $R_\lambda$. }
\label{fig:constraint1}
\end{figure}

In this paper, we use the X-ray peaks as the cluster fiducial center, and model the offsets between X-ray peaks and \RM~centers to characterize the \RM~centering distribution.

When \RM~misidentifies the galaxy at the cluster center, an offset between X-ray peaks and the \RM~centers is expected. On the other hand, when the \RM~centers are correct, the X-ray peaks may still be offset from them because of the different dynamics and relaxation timescales of gas and galaxies \citep[e.g., see studies about cluster state and X-ray galaxy offsets in][]{2015MNRAS.449..199M, 2017MNRAS.469.1414K, doi:10.1093/mnras/sty131}, as well as observational uncertainties in identifying the X-ray peaks \footnote{The photometric and astrometric uncertainties of galaxy positions are negligible in DES compared to X-ray \citealt{2017arXiv170801531D}}, but these offsets tend to be small, i.e., less than tens of \kpc. Therefore, we expect the well and mis-centered \RM~clusters to have different offset distributions to the X-ray peaks, and we model the \RM~and X-ray offset as a mixture of well-centered and mis-centered components, written as:
\begin{equation}
\begin{split}
P(x | \rho, \sigma, \tau) &=\rho \times 
P_\mathrm{cent}(x | \sigma) + (1-\rho) \times P_\mathrm{miscent}(x | \tau), \\
P_\mathrm{cent}(x | \sigma)& = \frac{1}{\sigma} \mathrm{exp}(- \frac{x}{\sigma}), \\
P_\mathrm{miscent}(x | \tau) &= \frac{x}{\tau^2} \mathrm{exp}(- \frac{x}{\tau}).\\
\label{eq:offset}
\end{split}
\end{equation}

In the above model, the peaked exponential distribution characterized by the parameter $\sigma$, $P_\mathrm{cent}(x | \sigma)$,  describes the X-ray offset distribution of the correctly centered clusters. The second component, $P_\mathrm{miscent}(x | \tau)$, a Gamma distribution of shape parameter 2 and characterized by a scale parameter $\tau$, describes the offset of the mis-centered clusters. This distribution has a heavy tail and the choice of this distribution is inspired by the Chandra SDSS sample having an extended offset distribution. The fraction of well-centered clusters are modeled by the $\rho$ parameter. In total, the models consists of three parameters $\rho$, $\sigma$ and $\tau$.  The \RM~and X-ray center offset, $x$, is computed as 
\begin{equation}
\begin{split}
x &=r_\mathrm{offset}/R_\lambda, \\
R_\lambda &=(\lambda/100)^{0.2}  h^{-1}\mpc,
\end{split}
\end{equation}
which scales the offset with a mild dependence on \RM~richness denoted by $\lambda$. 

Given the measurements of these offsets, $\{x_i\}$, the Bayesian posterior distribution of  $\rho$, $\sigma$ and $\tau$ is written as
\begin{equation}
\begin{split}
P(\rho, \sigma, \tau | \{x_i\}) &\propto P(\{x_i\}|\rho, \sigma, \tau) P(\rho, \sigma, \tau) \\
&= P(\rho, \sigma, \tau)\prod_{i}P(x_i|\rho, \sigma, \tau),
\end{split}
\end{equation}
where $P(\rho, \sigma, \tau)$ is the prior distribution of $\rho$, $\sigma$ and $\tau$ listed in Table~\ref{tbl:centering_params}, which is chosen to be flat and independent parameters. We sample the posterior distribution of $\rho$, $\sigma$ and $\tau$ using a Markov Chain Monte Carlo (MCMCc ) method.

\subsection{Model Constraints}
\label{sec:offset_constraint}

\begin{table}
\caption{Centering offset Parameter constraints (Equation~\ref{eq:offset}) for the Chandra DES and SDSS redMaPPer samples.}
\label{tbl:centering_params}
\vspace{1em}
\begin{tabular}{lllll}
\hline
 & $\rho$  & $\sigma$  & $\tau$\\
  \vspace{1em}
Prior & $[0.3, 1]$ & $[0.0001, 0.1]$ & $[0.08, 0.5]$ \\
  \hspace{0.5em} \\
\hline
\\
Chandra SDSS Posterior & $0.678\substack{+0.035 \\ -0.051}$ &  $0.0156\substack{+0.0026\\ -0.002}$  &  $0.179\substack{+0.021 \\ -0.021}$ \\
  \hspace{0.5em} \\
\hline
\\
Chandra DES Posterior & $0.835\substack{+0.112 \\ -0.075}$ &  $0.0443\substack{+0.0231\\ -0.0094}$  &  $0.166\substack{+0.111\\ -0.042}$ \\
  \hspace{0.5em} \\
\hline
 \\
\end{tabular}
\end{table}

The aforementioned X-ray \RM~offset model is constrained separately for the Chandra SDSS and DES \RM~samples. Table~\ref{tbl:centering_params} lists and Figure~\ref{fig:constraint1} shows the posterior constraints of the model parameters including the correctly-centered fraction $\rho$, and the mis-centered characteristic offset $\tau$, as well as the characteristic \RM~X-ray offset, $\sigma$. The SDSS sample yields higher precision because of the larger sample sizes. The fraction of well-centered clusters $\rho$ and the mis-centering offset $\tau$ for the mis-centered clusters are mildly different  from the DES \RM~sample which displays a hint of having a higher fraction of well-centered clusters at a $1.5\sigma$ significance level. For the well-centered clusters, the characteristic \RM~X-ray offset, $\sigma$, of the DES sample is larger than the respective parameter of the SDSS sample, reflecting the limited angular resolution of X-ray peak identification and the higher redshift range of the DES sample (and therefore lower physical separation resolution of the DES X-ray peak identification). 

The \RM~algorithm computes a centering probability, $P_\mathrm{cen}$, as an indicator of whether or not the selected central galaxy is the right choice. We do not find the values of $P_\mathrm{cen}$ to accurately reflect the centering statistics of the \RM~sample. Specifically, we study the dependence of the centering performance by separately constraining the centering model of the \RM~SDSS sample in two $P_\mathrm{cen}$ ranges, $\geq 0.9$ and $<0.9$ respectively. Clusters of $P_\mathrm{cen} \geq 0.9$ have centering fraction of $0.76\pm0.05$ while clusters of $P_\mathrm{cen}<0.9$ have a notably lower centering fraction of $0.49\pm0.09$. Although a larger  $P_\mathrm{cen}$ value does indicate a better centering performance, it does not reflect the real centering fractions at the face values.

We have tested the  dependence of the centering parameters through constraining the model with the SDSS sample in different ranges of richnesses, X-ray temperatures or luminosities, and for serendipitous vs targeted observations. We do not find significant differences in the centering parameter constraints. A larger set of X-ray observations would be needed to reveal any trends.

Interpretation of the results from this centering offset characterization depends on the adopted models. Different functional forms of the mis-centering offset -- a Rayleigh distribution for the mis-centered component, and a Gaussian distribution for the well-centered clusters -- have been attempted in previous studies \citep{2015MNRAS.454.2305S, 2016ApJS..224....1R, 2018MNRAS.480.2689H}. We have tested alternative models (listed in Table~\ref{tbl:model_comp}) for the mis-centering and centering distributions with the Chandra SDSS sample. As the common parameter shared across different models, the well-centered cluster fraction, $\rho$, is consistently constrained to be in the range of 64\% to 70\% for SDSS. 

We use the Bayesian deviance information criterion \citep[DIC,][]{citeulike:105949} to compare the fitness of the models listed in Table~\ref{tbl:model_comp}. We compute the Bayesian DIC values \citep[DIC,][]{citeulike:105949}  by sampling the posterior constraints of the alternative and nominal models presented in the paper. 
DIC is computed as 
\begin{equation}
DIC= -2 \overline{\mathrm{log} (p(\{x_i\}| \theta )}+2 \mathrm{log} (p(\{x_i\}|\bar{\theta}).
\end{equation}
$\overline{\mathrm{log} (p(\{x_i\}| \theta )}$ is the probability of the observed  offsets averaged over the posterior mis-centering model, and $\mathrm{log} (p(\{x_i\}| \bar \theta )$ is the probability of the observed  offsets given the best-fitting mis-centering model. Lower DIC values indicate better fitting of the model, and a DIC difference larger than 2 is considered significant.  This DIC comparison strongly favors the nominal model described in this section.

\begin{table*}
\caption{Alternative offset models and constraints derived with the Chandra SDSS redMaPPer sample, and DIC comparisons to the nominal model in this paper. }
\label{tbl:model_comp}
\vspace{1em}
\begin{tabular}{lllccc}
\hline
 ~~~ Name ~~~ &  ~~~ Model ~~~ &  ~~~Constraints ~~~&   ~~~ DIC -  DIC$_\mathrm{nominal}$  \\
  \vspace{1em}
    \hspace{0.5em} \\
\hline
\\
Gaussian & $P(x | \rho, \sigma, \tau) =\rho  
P_\mathrm{cent} + (1-\rho)P_\mathrm{miscent} $ & $\rho =0.64\substack{+0.05 \\ -0.05}$ & 19.0 \\
& $P_\mathrm{cent}(x | \sigma) = \frac{2}{\sqrt{2\pi}\sigma} \mathrm{exp}(- \frac{x^2}{2\sigma^2}) $ & $\sigma =0.0177\substack{+0.0025 \\ -0.0020}$ \\
 & $P_\mathrm{miscent}(x | \tau) = \frac{x}{\tau^2} \mathrm{exp}(- \frac{x}{\tau}).$ & $\tau =0.161\substack{+0.021 \\ -0.016}$  \\
   \hspace{0.5em} \\
\hline
\\
Rayleigh & $P(x | \rho, \sigma, \tau) =\rho  
P_\mathrm{cent} + (1-\rho)P_\mathrm{miscent} $ & $\rho =0.70\substack{+0.05 \\ -0.04}$ & 4.4 \\
& $P_\mathrm{cent}(x | \sigma) = \frac{1}{\sigma} \mathrm{exp}(- \frac{x}{\sigma}) $ & $\sigma =0.0185\substack{+0.0025 \\ -0.0023}$ \\
 & $P_\mathrm{miscent}(x | \tau) = \frac{x}{\tau^2} \mathrm{exp}(- \frac{x^2}{2\tau^2}).$ & $\tau =0.323\substack{+0.029 \\ -0.024}$  \\
  \hspace{0.5em} \\
\hline
\\
Full Gamma & $P(x | \rho, \sigma, \tau, k) =\rho  
P_\mathrm{cent} + (1-\rho)P_\mathrm{miscent} $ & $\rho =0.64\substack{+0.07 \\ -0.05}$ & 1.85 \\
  (Four Parameters) & $P_\mathrm{cent}(x | \sigma) = \frac{1}{\sigma} \mathrm{exp}(- \frac{x}{\sigma}) $ & $\sigma =0.015\substack{+0.0027 \\ -0.003}$ \\
 & $P_\mathrm{miscent}(x | \tau, k) = \frac{x^{k-1}}{\Gamma(k) \tau^k} \mathrm{exp}(- \frac{x}{\tau})$ & $\tau =0.21\substack{+0.07 \\ -0.05}$  \\
  & & $k=1.0\substack{+0.83 \\ -0.0}$ \\
  \hspace{0.5em} \\
\hline
\\
Cauchy & $P(x | \rho, \sigma, \tau) =\rho  
P_\mathrm{cent} + (1-\rho)P_\mathrm{miscent} $ & $\rho =0.645\substack{+0.05 \\ -0.05}$ & 14.2\\
& $P_\mathrm{cent}(x | \sigma) = \frac{1}{\sigma} \mathrm{exp}(- \frac{x}{\sigma}) $ & $\sigma =0.014\substack{+0.0028 \\ -0.0022}$ \\
 & $P_\mathrm{miscent}(x | \tau) = \frac{x\tau}{(x^2+\tau^2)^{1.5}}$ & $\tau =0.14\substack{+0.04 \\ -0.03}$  \\
  \hspace{0.5em} \\
\hline
\\
\end{tabular}
\end{table*}

Notably, when adopting a Rayleigh distribution for the mis-centered component, our SDSS posterior values on the well-centered cluster fraction, $\rho$, and the mis-centered characteristic offset, $\tau$ is highly consistent with the previous study in \cite{2018MNRAS.480.2689H}  which adopts a similar model and is based on a similar SDSS \RM~catalog. The posterior precision of the parameters has improved significantly in our analysis. 

\subsection{Model Validation}

\begin{figure}
 \centering
 \includegraphics[width=0.95\linewidth]{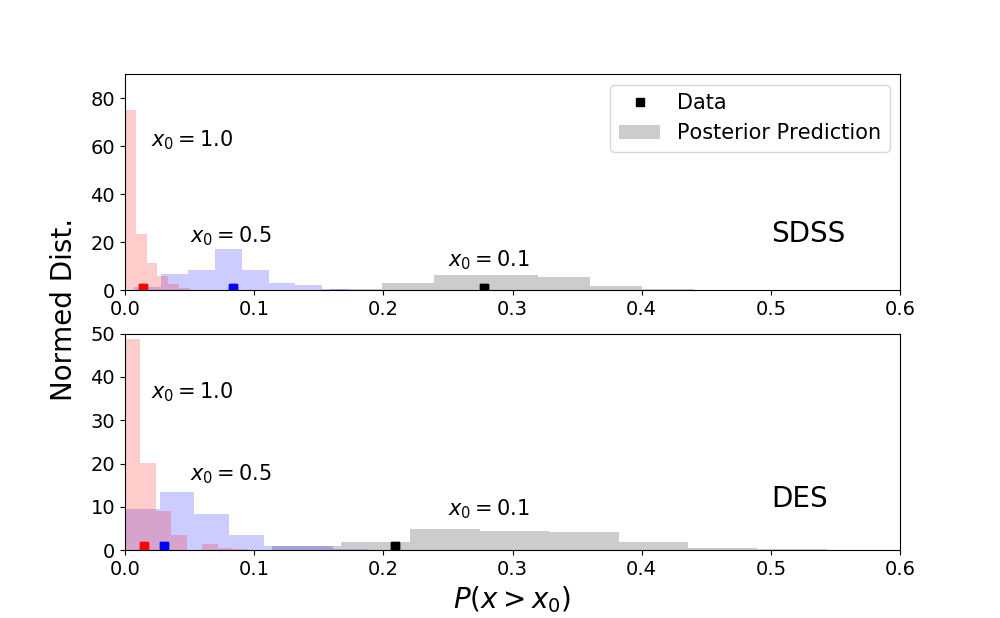}
 \caption{Posterior predictive check of the centering offset model. We show the predictions on the fraction of clusters in different offset ranges ($x>0.1$, $x>0.5$ and $x>1.0$) from the offset model sampling the posterior constraints (histograms). These predictions agree with the measurements from data (solid squares) for both the SDSS and DES models and data.}
 \label{fig:pp_fraction}
\end{figure}

We check the model goodness-of-fit with a posterior prediction test. Specifically, we compare the fractions of clusters in an offset range from the data to the predictions from the constrained models. The procedures are as the follows.

\begin{enumerate}
\item From the measurements of $x=r_\mathrm{offset}/R_\lambda$, record the number of clusters with offsets larger than a comparison value $x_0$. 
\item Take one set of model parameters, $\rho$, $\sigma$ and $\tau$ from the MCMC posterior constraints, denoted as $\rho_i$, $\sigma_i$ and $\tau_i$. Randomly draw a set of centering offsets, $\{x_{ij}\}$, from the offset model (Equation~\ref{eq:offset}) with the above set of posterior model parameters. The number of random draws should match the size of the X-ray-\RM~ sample being tested. 
\item With the above set of centering offsets sampling, $\{x_{ij}\}$, record the number of offsets larger than a comparison value $x_0$, $N(x_{ij}>x_0)$.
\item Repeat the process for each set of $\rho$, $\sigma$ and $\tau$ values from the MCMC posterior chain and acquire the distribution of $N$. This is the posterior prediction on the number of clusters with offsets larger than the comparison value $x_0$.
\item Compare the number from data to this posterior prediction. We expect a two-sided $P$-value, defined as the minimum of the fractions of the posterior predictions above and below the data, to be larger than 0.025.
\end{enumerate}

We use the above process to evaluate the goodness of the model at offsets larger than 0.1, 0.5 and 1 $R_\lambda$.  Figure~\ref{fig:pp_fraction} shows the posterior predictive distribution of these offset ranges. For both the SDSS and DES \RM~samples,  the prediction from the model and its respective model constraints well match the measurements from the data in these small, medium and large offset ranges, with the two-sided $P$-values being 0.35, 0.37 and 0.18 for SDSS and 0.11, 0.25 and 0.17 for DES.

\section{Mis-centering Impact on Richness Scaling Relation}
\label{sec:lambda_model}

When a cluster is mis-centered, the \RM~richness estimation may become biased, and the bias depends on the mis-centering offset. In this section, we use the \RM~catalog itself to constrain a model that describes the bias of $\lambda$ upon a mis-centering offset.

As mentioned in Section~\ref{sec:center}, the~\RM~algorithm selects the five most probable central galaxies and stores the $\lambda$ estimations computed at each of the centers. The default ~\RM~center is chosen as the one with the highest centering probability. We make use of this information to construct a richness  shift {\it VS} offset model. 

Assuming that there existed a~\RM~catalog with the most probable centers always being the correct ones, the $\lambda$ estimations computed at the other four centers will be affected by the mis-centering effect. The $\lambda$ bias between the real center and each of the four remaining center candidates, and the positional offsets between them, can be used  to constrain a $\lambda$ {\it vs.} centering offset model.

We use the lambda offsets between the 2nd and the 1st (\RM~default center) most likely centers, and the distance offset between them, to model the richness {\it VS} centering offset (Section~\ref{sec:lambda_model_constraints}). The model is further validated with X-ray data (Section~\ref{sec:xray-validation}).

\subsection{Model and Model Constraints}
\label{sec:lambda_model_constraints}

\begin{figure*}
\includegraphics[width=0.85\linewidth]{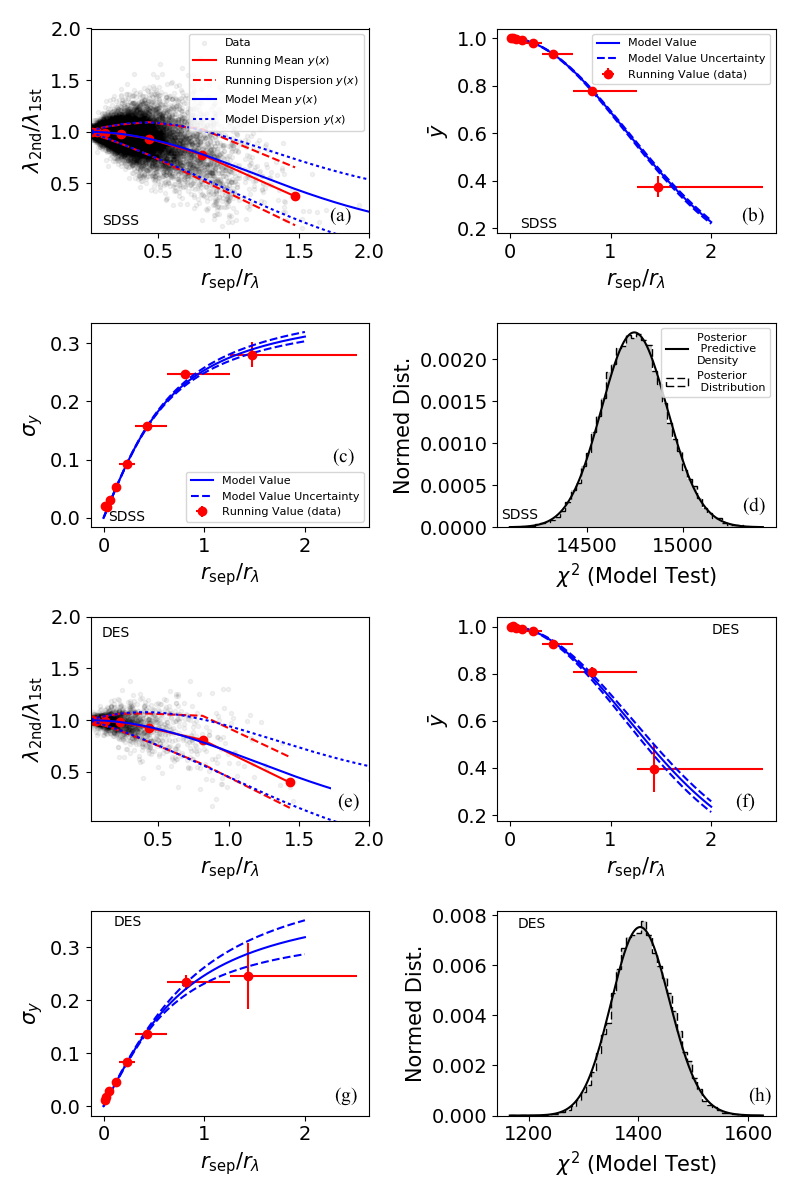}
\caption{The richness  VS offset distributions and models. The cluster richness tends to be biased lower by mis-centering (a, e), and the mean of the bias can be characterized by a Gaussian function (b, f). The biases have large dispersions (c, g), which can be further characterized by an $\mathrm{arctan}$ function (c, g). Posterior predictive checks show that the models are tightly constrained from the data and fit the data well (d, h).}
\label{fig:lambda_center}
\end{figure*}

\begin{figure}
\includegraphics[width=0.95\linewidth]{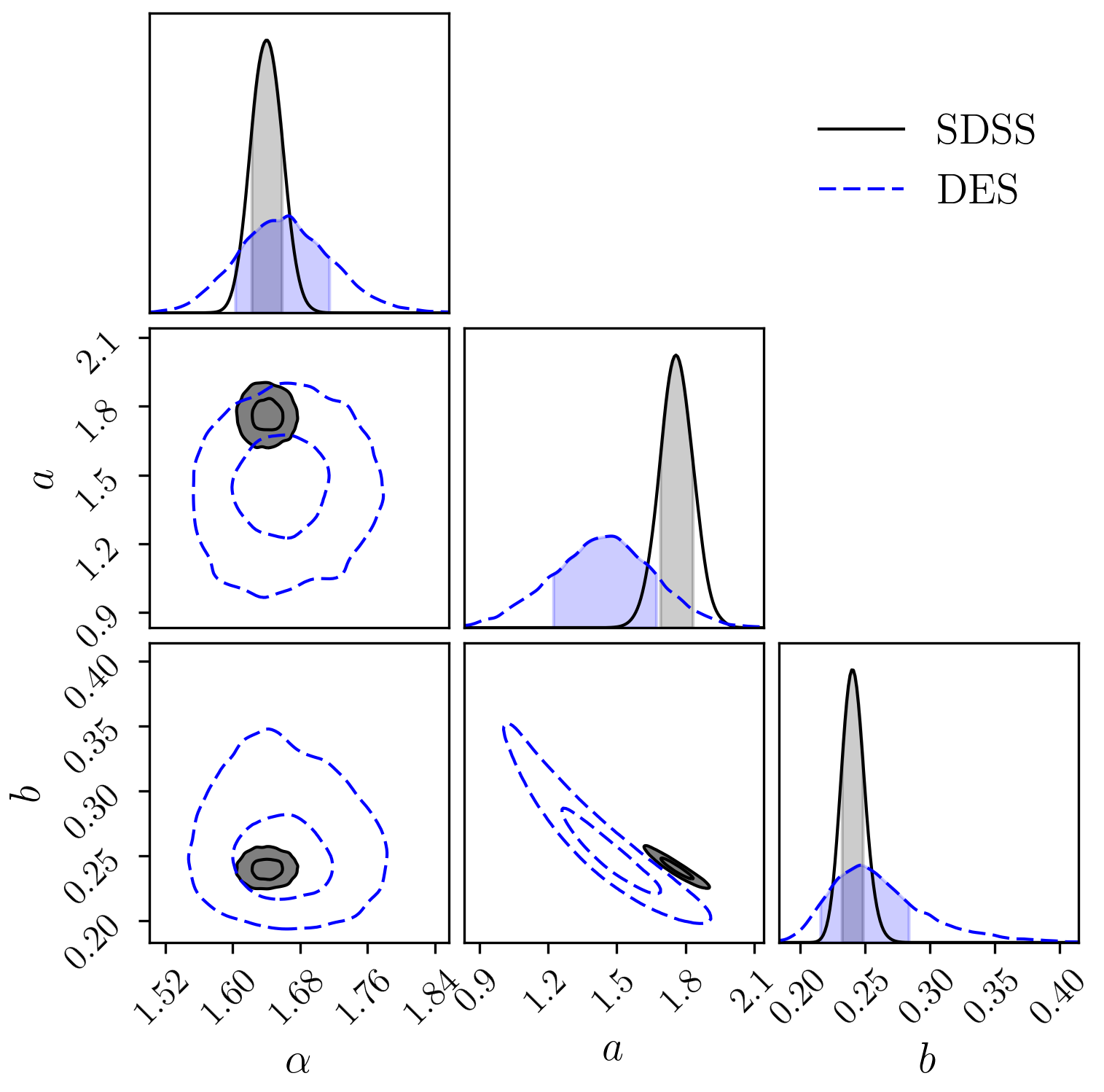}
\caption{Parameter constraints for the lambda offset VS centering offset model. See Section 4.2 for details.}
\label{fig:lambda_center_mcmc}
\end{figure}

\begin{table}
\caption{Parameter constraints for the lambda offset VS centering offset model as in $\bar{y}(x)=\mathrm{exp}(-x^2/\sigma^2)$  and $\sigma_y (x)=a\times\mathrm{arctan}(bx)$. }
\label{tbl:mcmc}
\vspace{1em}
\begin{tabular}{lllll}
\hline
\vspace{1em} \\
 & SDSS $\alpha$  & SDSS $b$  & SDSS $a$\\
  \vspace{1em}
Prior &  (0, 10) & (0, 10) & (0, 10)\\
Posterior & 1.64$\pm$0.02& 1.76$\pm$0.07 & 0.241$\pm$0.008\\
\hline
\vspace{1em} \\
 & DES $\alpha$  & DES $b$  & DES $a$\\
   \vspace{1em}
Prior &  (0, 10) & (0, 10) & (0, 10)\\
Posterior & 1.66$\pm$0.06& 1.43$\pm$0.22 & 0.26$\pm$0.04\\
\hline
\vspace{1em}
\end{tabular}
\end{table}

We quantify the fractional shift of $\lambda$, $y= \lambda_\mathrm{miscentered}/\lambda_\mathrm{true}$ due to mis\-centering. This shift is quantified as being dependent on the scaled mis-centering offset in terms of $r_\lambda$. Upon trial of different analytical forms, we model the probabilistic distribution of $y$ as a Gaussian distribution:
\begin{equation}
y\sim\mathcal{N}(\bar{y}(x), \sigma_y (x)),
\end{equation} 
with the mean and the dispersion, $\bar{y}( x)$ and $\sigma_{y} ( x)$, both depending on $x$. The mean is positive, decreases with larger offsets starting from 1, and asymptotically reaches 0 for large offsets (Figure~\ref{fig:lambda_center}). The dispersion is positive, increases with larger offsets starting from 0, and asymptotically reaches a constant for large offsets (Figure~\ref{fig:lambda_center}). Specifically, we choose $\bar{y}( x)$ to be a Gaussian function between $y$ and $x$, $\bar{y}(x)=\mathrm{exp}(-x^2/\alpha^2)$ with $\alpha$ being a model parameter. $\sigma_y (x)$ is chosen as $\sigma_y (x)=a\times\mathrm{arctan}(bx)$ and $a$ and $b$ are model parameters.

The process of evaluating the above model goes as follows:
\begin{enumerate}
\item We compute the separation between the 1st and the 2nd most probable centers, $r_\mathrm{sep}$, assuming the cluster photometric redshift from the \RM~algorithm for both of the galaxies. We scale $r_\mathrm{sep}$ as $r_\mathrm{sep}/r_\lambda$ to become quantity $x$.  
\item We compute the relative $\lambda$ offset between the 1st and the 2nd most probable centers as $y= \lambda_\mathrm{2nd}/\lambda_\mathrm{1st}$.
\item We repeat the above process for each cluster $i$ in the \RM~catalog, acquiring a measurement data set  $\{x_i\}$ and $\{y_i\}$. This data set is shown in Figure~\ref{fig:lambda_center}.
\item With the above measurement data set, we constrain the model parameters of $y$ with the following likelihood: $\mathcal{L}=\sum_{i, x_i<0.1}\big( -\frac{[y_i-\bar{y}(x_i)]^2}{2\sigma^2_y(x_i)}-\mathrm{ln}[\sigma_y(x_i)]  \big) $. The likelihood is sampled with a MCMC algorithm.
\item Note that only the data points with $x_i > 0.1$ are used in the fitting process. This helps eliminate over-fitting at small $x$, and improves the fitting results at large $x$. 
\end{enumerate} 

The same measurements and modeling processes are performed for the SDSS and DES \RM~catalogs separately. Figure~\ref{fig:lambda_center} shows the data points as well as the best fitted models in the first two columns. For comparison, the red lines/points are the data running averages and running dispersions in the $x$ bins (bin widths indicated by the $x$ error bars in the second column).
The MCMC posterior constraints are shown in Figure~\ref{fig:lambda_center_mcmc} and listed in Table~\ref{tbl:mcmc}. $\alpha$ does not appear to be covariant with $a$ or $b$, but $a$ and $b$ appear to be highly covariant.

Overall, the impact of mis-centering on cluster richness is mild  with a low scatter at small mis-centering offset, but grows  with a larger offset, reaching a bias ratio $\lambda_\mathrm{miscentered}/\lambda_\mathrm{true}$ of 0.5 at  $r_\mathrm{sep}/R_{\lambda} \sim 1.40 $.


\subsection{Model Validation with X-ray Centers}
\label{sec:xray-validation}

\begin{figure*}
\includegraphics[width=0.95\linewidth]{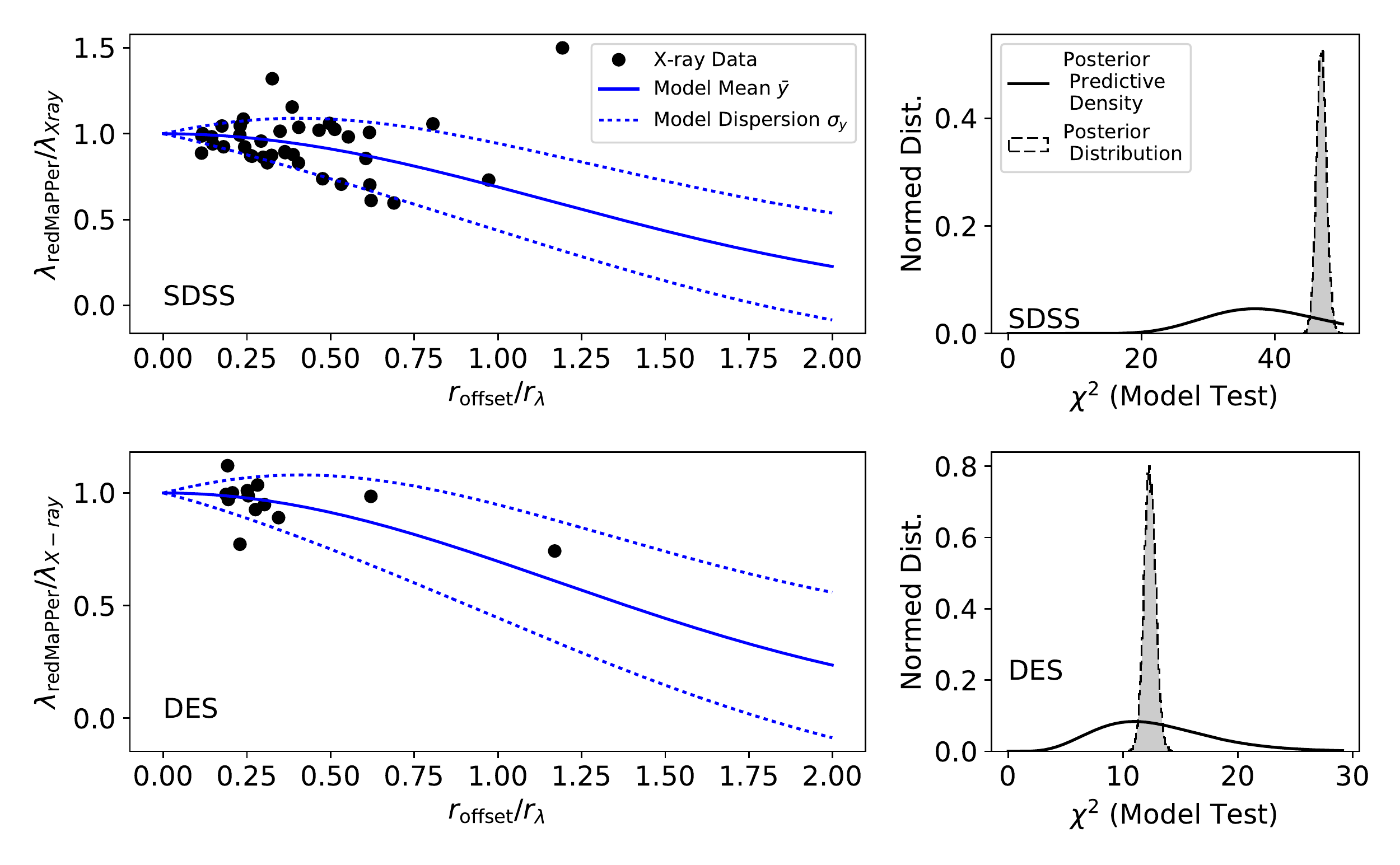}
\caption{The shift in richness when the richness is estimated at the X-ray centers rather than the redMaPPer centers ($1st$ column). The lambda VS centering offset models derived from the redMaPPer catalog are shown as blue solid and dashed lines. Posterior predictive checks ($2nd$ column) show that the richness shifts estimated at the X-ray centers adhere to the models derived from the redMaPPer catalog (see Section~\ref{sec:xray-validation} for details).}
\label{fig:chi2_xray_des}
\end{figure*}

As a test of the richness offset model, we make use of the Chandra-\RM~samples (Section~\ref{sec:chandra-redmapper}) and compare the X-ray peak \RM~centered $\lambda$ to the prediction from the model.

We rerun the redMaPPer $\lambda$ algorithm with the X-ray peaks as the cluster centers. The procedures are equivalent to the original $\lambda$ estimation with the exception of a  "percolation" process (Rykoff 2014), which re-evaluates $\lambda$ upon masking neighboring \RM~clusters. The $\lambda$ estimations on X-ray peaks do not go through the "percolation" process as the run does not consider \RM~clusters not present in the X-ray sample. In this test, to ensure that the "percolation" process is negligible, we remove clusters whose $\lambda$ changed by 10\% in the initial \RM~percolation process\footnote{Two from each of the SDSS and DES Y1 samples.}.

We calculate the $\lambda$ offsets VS the distance offsets between the X-ray peaks and \RM~centers. The $\lambda$ offsets are calculated as $y_x= \lambda_\mathrm{RM}/\lambda_\mathrm{xray}$, and the corresponding X-ray and \RM~mis-centering offsets as $x_x=r_\mathrm{sep}/r_\lambda$, where $r_\lambda$ is evaluated with the X-ray centered $\lambda$. Clusters of centering offsets $x_x$ less than 0.1 are considered well-centered (a similar cut is applied when deriving the model in Section~\ref{sec:lambda_model_constraints}) and do not enter the test. In Figure~\ref{fig:chi2_xray_des}, we show the derived $\lambda$ and centering offsets from the X-ray observations. The constrained models in Section~\ref{sec:lambda_model_constraints} appear to be qualitatively consistent with these offsets.

To quantify the fitness of the model from the previous section, we compute the following $\chi^2$ discrepancy given model parameters and observations:
\begin{equation}
\chi_n^2(\alpha, a, b)=\sum^n_{i=1} [ \frac{y_{x,i} -\bar{y}(x_{x,i})}{\sigma_y(x_{x,i})} ]^2.
\end{equation}
As described in the previous section, $\alpha$ is the parameter of the $\bar{y}(x)$ function, and  $a$ and $b$ are the parameters of the $\sigma_y(x)$ function.
If the model quantitatively describes cluster mis-centering correctly, and the values of $\alpha$, $a$ and $b$ are accurate, $\chi_n^2(\alpha, a, b)$ will appear to be drawn from a chi-squared distribution, $\chi^2(n)$, with the degree of freedom, $n$,  matching the number of $\{y_{x}, x_{x}\}$ observations. This chi-squared distribution is known as the posterior predictive density for the $\chi^2$ discrepancy.

We perform Bayesian posterior predictive assessment on the fitness of the model following the process in \cite{10.2307/24306036, 10.2307/2242219}. The process includes calculating a posterior predictive $p-$value (PPP value) which is the classical $p-$value averaged over the posterior model parameter distribution. With the posterior distribution of $\alpha$, $a$ and $b$ sampled with the Monte Carlo Markov Chain (MCMC) method, the procedure goes as follows:
\begin{enumerate}
\item Take one set of $\alpha$, $a$ and $b$ from the MCMC posterior constraints, donated as $\alpha_j$, $a_j$ and $b_j$.
\item Calculate the $\chi^2$ discrepancy for $\alpha_j$, $a_j$ and $b_j$, denoted as $\chi^2_j$.
\item Repeat the process for each set of $\alpha$, $a$ and $b$ values from the MCMC posterior chain. 
\item For each $\chi^2_j$, randomly draw a value, $q_j$, from a standard $\chi^2(n)$ distribution. For the posterior set of $\{\chi^2_j, q_j\}$, record the fraction of $\chi^2_j \geq q_j$ as $Pb_1$, and the frequency of $\chi^2_j \leq q_j$ as $Pb_2$. We use a two sided $p-$value definition, that $Pb=\mathrm{min}(Pb_1, Pb_2)$ as the PPP value.
\end{enumerate}

We compute the posterior $\chi^2$ discrepancy given X-ray $\{y_x, x_x\}$ observations shown in Figure~\ref{fig:chi2_xray_des}. The posterior $\chi^2$ discrepancy values are expected to occupy a highly-probable interval of a chi-squared distribution $\chi^2(n)$  (posterior predictive density), with the degree of freedom $n$ matching the number of  $\{y_x, x_x\}$ observations from X-ray peak centers. This appears to be the case for both the SDSS and DES X-ray \RM~sample. The Bayesian posterior predictive P values (PPP values) are 0.18 and 0.50 respectively for the SDSS and DES X-ray \RM~samples. Both of the PPP values are above a 0.025 model rejection threshold, indicating consistency between the constrained models and the offsets derived with X-ray peak centers.

We also compute the posterior $\chi^2$ discrepancy for the original $\{y_i\}$, $\{x_i\}$ observations that are used to constrain the model. The distributions of $\{\chi^2_j\}$ are shown as the gray shaded histogram in Figure~\ref{fig:lambda_center}, along with the probability density of a chi-squared distribution, $\chi^2(n)$, for comparison. The distribution of $\{\chi^2_j\}$ is in good accordance with the expected $\chi^2(n)$ distribution, for both the SDSS and DES \RM~samples, indicating the goodness of the fit and tightness of the model parameter constraints. The PPP values for the SDSS and DES samples are 0.50 and 0.49 respectively, matching the expectation of a well-posited model.

Note that when deriving the model, the available~\RM~catalogs already have imperfect center selections. According to the previous section, the majority ($\sim$70\%) of the clusters in the~\RM~catalog are correctly selected. We have attempted to select clusters with a higher centering probability (using the \RM~$P_\mathrm{cen}$ quantity), but the samples selected on $P_\mathrm{cen}$  display a hint of performing slightly worse in the validation test with X-ray data.  Because of a concern that this $P_\mathrm{cen}$ selection may have biased the cluster sample, and that the selection still cannot ensure a 100\% well centered subsample, we do not apply the  selection in this paper and emphasize on our derived model passing the validation test with X-ray data. In the future, it would be desirable to quantify the richness offset model with X-ray centered richness when larger X-ray redMaPPer clusters become available.

\section{Implications for DES Cluster Cosmology}
\label{sec:cosmo}

\subsection{Mis-Centering Model in DES Cluster Weak Lensing Analysis}

\begin{figure}
 \centering
  \includegraphics[width=0.95\linewidth]{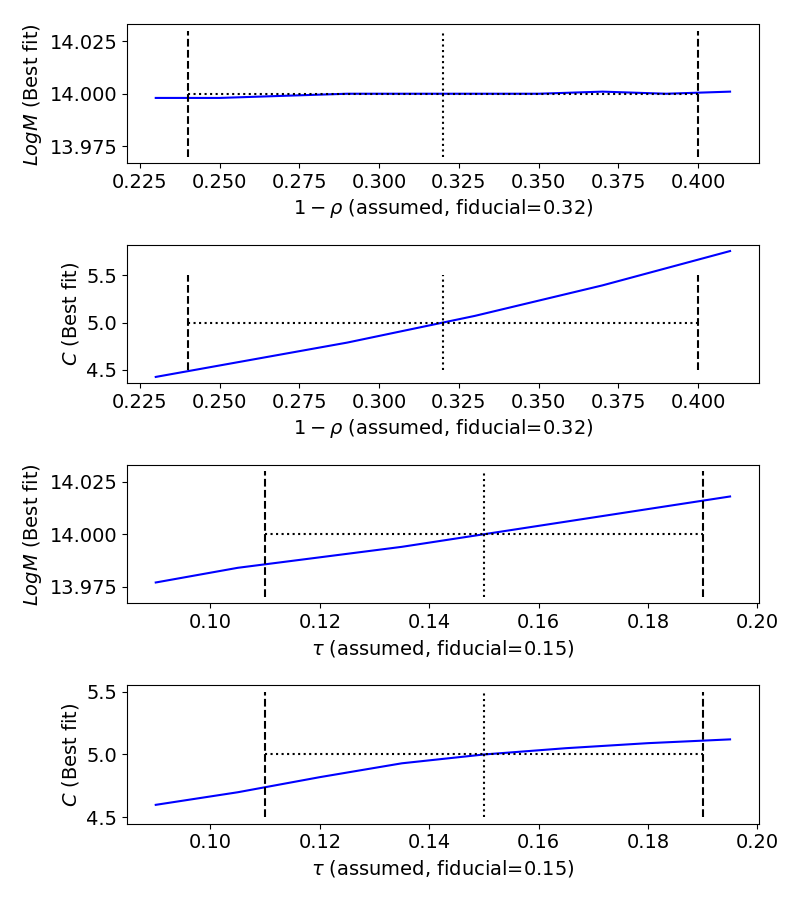}
 \caption{We explore how inaccurate knowledge of the mis-centering model parameters affects the accuracy of cluster mass and concentration estimations in cluster lensing studies. A fiducial mass profile is created using the prescription in McClintock et al. 2018 with a fiducial set of mis-centering parameters. Through comparing to the fiducial mass profile, the best fit mass and concentrations are estimated for different assumed values of the mis-centering model parameters, $\rho$ and $\tau$. We find that the mass estimation is robust under inaccurate assumptions of $\rho$, but susceptible to inaccuracy in $\tau$. The concentration parameter, on the contrary, is more susceptible to the inaccuracy of $\rho$ than $\tau$. The vertical lines indicate parameter ranges comparable to those in McClintock et al. 2018.}
\label{fig:m_c_vary}
\end{figure}

In DES stacked cluster lensing studies, the cluster lensing signals are fitted to an analytic model to determine cluster mass and concentration. To summarize, \cite{McClintock2018} adopts the following method to correct for mis-centering:
\begin{equation}
\Sigma(r|M, c) = \rho \Sigma_\mathrm{cent}(r|M, c) + (1-\rho)\Sigma_\mathrm{miscent}(r|M, c)
\end{equation}
In the above equations, $\Sigma(r|M, c)$ is the cluster mass profile model with mass $M$ and concentration $c$. $\Sigma_\mathrm{cent}(r|M, c)$ denotes the mass profile model for well-centered clusters, while $\Sigma_\mathrm{miscent}(r|M, c)$ is the mass profile model for mis-centered clusters. The mis-centered profile
is averaged over the angle, $\theta$, and magnitude, $R$, of the radial vector to the correct center,
\begin{equation}
\begin{split}
\Sigma_\mathrm{miscent}(\vec{r}|M, c)=&\Sigma_\mathrm{cent}(\vec{r}+\vec{R}|M, c),\\
\Sigma_\mathrm{miscent}(r|M, c)= & \frac{1}{2\pi}\int \mathrm{d} \theta \mathrm{d} R P_\mathrm{miscent}(R) \\ & \Sigma_\mathrm{cent} (\sqrt{r^2+R^2+2rR\mathrm{cos}(\theta)}|M, c).
\end{split}
\end{equation}
The distribution for the magnitude of the radial offset $R$,  is $P_\mathrm{miscent}(R)$, which is described by a parameter $\tau$.  The model does not account for the offset between central galaxies and X-ray peaks because halos are assumed to be centered on a massive dark matter substructure hosting a central galaxy.
Based on the analyses carried out in this study and a companion SDSS \RM~centering study of a complete sample with Swift observations in \cite{AvdLinden},  \cite{McClintock2018}  adopted the prior  $\rho=0.75\pm0.08$ and $\tau=0.17\pm0.04$. These values are consistent with the Chandra DES constraints presented in Section~\ref{sec:center}, but also encompass the results of the SDSS samples presented in this paper and in \cite{AvdLinden}.

\subsection{Sensitivity of Cluster Mass Estimation to the Mis-centering Model}

We determine the sensitivity of the mass calibration to variations in the values of the mis-centering parameters.  To do so, we create a fiducial mass profile and analyze how much the measured masses deviate from the truth if the mis-centering model is inaccurate. 
Following the recipe in \cite{McClintock2018}, the fiducial mass profile model combines a NFW profile and a two-halo model of $M_{200m}=$ $10^{14} $ M$_\odot$, concentration of 5 and $R_\lambda = 1$ Mpc. We compute the mis-centered lensing signal by adopting fiducial values $\rho=0.68$ and $\tau=0.15$.  We fit this synthetic weak lensing data with a minimum $\chi^2$ method assuming a range of
values for both $\rho$ and $\tau$, and measure the bias of the best fit mass and concentration as a function of these two parameters. The fitting process is restricted to the 0.2 to 30 Mpc radius range as in the DES Y1 weak lensing study \citep{McClintock2018} and the profile measurement uncertainty is assumed to be due to shape nose only, and therefore scales with radius as $r^{-1}$.

Figure~\ref{fig:m_c_vary} shows the best fit mass and concentration parameters as a function of $\rho$ and $\tau$. We find that the best-fit mass is insensitive to the assumed $\rho$ value, whereas the recovered concentration is biased. Allowing the concentration parameter to vary effectively decouples the recovered mass from $\rho$. By contrast, variations in $\tau$ have a non-negligible impact on the best-fit mass. Uncertainties in $\tau$ at a level of $\pm0.04$, comparable to the constraint in this paper, results in a mass uncertainty of $\pm$0.015~dex\footnote{0.015 dex means $\delta \mathrm{log} M_{200m} = 0.015$}. 

These results are in qualitative agreement with the results of \citet{McClintock2018}, though computed with two important methodological differences.  Specifically: 1) \citet{McClintock2018} has considered more systematic effects other than mis-centering, including using a semi-analytic covariance matrix \citep{2015MNRAS.449.4264G}, and accounting for boost-factor corrections  \citep{Varga:2018}.  These changes will affect the relative weighting of radial inner to outer scales, thereby impacting the sensitivity of the mass posteriors to the mis-centering parameters.  2) When constraining the richness-mass scaling relation parameters, \citet{McClintock2018} treats the mis-centering parameters for each richness and redshift bin as independent, which reduces the relative importance of mis-centering in their analysis.  Together, these differences reduce the sensitivity of the scaling relation amplitude from the 0.015 dex we estimated here to 0.78\%, as quoted in \citet{McClintock2018}.  Nevertheless, it is clear from Figure 10 in \citet{McClintock2018} that the mass posterior in a single bin is largely insensitive to $\rho$, but is degenerate with $\tau$, as illustrated in our toy model analysis above.

These conclusions, however, rely on the assumption that the cluster mass profile is not correlated with the cluster mis-centering effect in optical data. Future cluster lensing analysis may wish to further investigate this assumption, e.g., through examining the cluster mass distribution in X-ray selected clusters (Das et al., private communications).

\subsection{Cluster Abundance}

The $\lambda$ offset caused by cluster mis-centering introduces bias and scatter into the lambda-mass scaling relation. We study the scatter increase with a test based on a N-body dark matter simulation \citep[][]{2016NewA...42...49H}. Richnesses are prescribed to each of the simulation dark matter halos following the richness-mass relation in \cite{2015MNRAS.454.2305S}, with a richness scatter, $\sigma_{\mathrm{ln}\lambda| M}$, of 25\%. We perturb the assigned richnesses with the Chandra SDSS offset model presented in Section~\ref{sec:center} and the richness bias model presented in this section, and find the richness scatter, $\sigma_{\mathrm{ln}\lambda| M}$ to increase by 2\%.

The bias and scatter manifest themselves in the number count of clusters selected by $\lambda$, which is a fundamental input to cluster abundance cosmology. Mis-centering tends to lower the richness estimation and the numbers of clusters above a richness threshold selected by the mis-centered richnesses would be lower than those selected by richnesses without mis-centering. The average mass of the clusters selected by the mis-centered richnesses tend to be higher. Testing with simulations shows that the numbers of clusters selected by the mis-centered richnesses is lower by $\sim 2\%$, and the average cluster masses increase by $\sim0.5\%$. \cite{Costanzi2018} estimates these shift to have negligible effects for SDSS and DES Y1 cluster cosmology constraints since they are are significantly smaller than the remaining systematic uncertainties of cluster abundance and mass estimations. Nevertheless, \cite{Costanzi2018, DES2018} corrected the data vectors with the factors listed above to account for the mis-centering effect. Future cluster cosmology analysis in the coming years of DES and LSST may wish to explicitly incorporate the mis-centering richness relations as the mis-centering effect on cluster abundance becomes more substantial compared to the statistical uncertainty.

\section{Centering study with XMM data}
\label{sec:xmm}

During the preparation of this paper, an additional sample of \RM~clusters (SDSS and DES) with archival X-ray observations from the XMM-Newton space telescope became available. As the centering analyses in this paper are optimized for Chandra observations, especially the centering offset model for well-centered clusters is optimized for Chandra PSFs, we do not attempt to combine the XMM and Chandra observations in the modeling processes. Rather, we use the XMM defined X-ray centers to explore the robustness of the fits presented in Section~\ref{sec:center}. The sample selection and the XMM data analysis methods are described in Section~\ref{XMMdata} and Section~\ref{XMMChandra} describes the centering comparison results from XMM.

\subsection{X-ray Data Processing}
\label{XMMdata}

The \RM--XMM joint samples (SDSS and DES) were constructed as follows. First, the \RM~ centroids were compared to the aim points of observations in the XMM public archive\footnote{The archive match used in this analysis was carried out on August 2018.}. \RM~ clusters with centroids falling outside 13$^{\prime}$ of an aim point were excluded from the samples. Second, the mean and median XMM exposure time was determined within a 10$^{\prime\prime}$ radius of the \RM~ centroid. For this we used exposure maps produced by the XMM Cluster Survey \citep[XCS,][]{XCS}. Any \RM~ clusters with mean exposure times of $<3$~ks, and/or median exposure times of $<1.5$~ks, were excluded from the samples. We note that the median filter was necessary because some clusters straddle both active and inactive regions of the field of view (FOV), e.g. those lying close to the FOV edge. If a given cluster was observed multiple times by XMM, only the observation with the longest exposure time, at the \RM~centroid, was used in subsequent analyses. Third, the remaining \RM~cluster centroids were compared to the list of extended sources detected using the XCS Automated Pipeline Algorithm (XAPA). Any \RM~ clusters  lying further than $2~h^{-1}$~Mpc (assuming the \RM~redshift) of such a source were excluded from the samples.  At this stage, the SDSS and DES \RM--XCS samples comprised of 356 and 282 clusters respectively. 




The X-ray peaks for the clusters in the \RM--XMM joint samples were then determined using a method that closely follows that used for the Chandra analyses, as described in Section~\ref{sec:chandra-redmapper}. An initial peak location was found in the respective merged (PN, MOS1, MOS2) {\em XMM} image, after smoothing with a $\sigma=50h^{-1}$~kpc Gaussian (assuming the \RM~redshift). As with the {\em Chandra} analysis, other sources (i.e. those assumed not to be associated with the \RM~ cluster) are masked out before the smoothing takes place. For this, we use the XAPA source catalogue of point-like sources. If there are multiple XAPA extended source within the image, then only the one closest to the \RM~ centroid is left unmasked. The peak location is then the brightest pixel in the masked, smoothed image, within a radius 1.5$\times$R$_{\lambda}$ of the RM position. 

The initial peak selection was occasionally erroneous. For example when there was a very bright point source in the XMM FOV. Such sources ``bleed'' into the surrounding region of the detector, meaning that the default point source mask size was not large enough remove them completely. Such cases were easily identified by eye using SDSS (or DES) and XMM ``postage stamp'' images. Most of these cases could be corrected by adjusting the size of the point source mask, and then re-running the peak finding script. However, for some, the point source ``bleeding'' was so pronounced that the respective cluster had to be removed from the \RM--XMM joint sample. Another reason for the initial peak being erroneous was the mis-percolation issue described in Section~\ref{sec:chandra-redmapper}. The mis-percolation cases were also identified using eye-ball checks. For these it was necessary to adjust the extended source mask, so that the closest source was now masked, but the second closest was not.


Once the second run of peak finding has been completed (and the new peak positions have been confirmed by eye), the remaining clusters were eye-balled again. At this stage, more \RM~clusters were removed from the sample: {\it i)} those where the XAPA extended source is clearly not associated with the \RM~ cluster (i.e. it is a foreground background cluster in projection), and {\it ii)} those where the \RM~ central galaxy falls in an XMM chip gap. After the various cuts described above, the SDSS--XCS and DES--XCS samples contained 248 and 109 sources respectively. 


\begin{figure}
 \centering
 \includegraphics[width=0.95\linewidth]{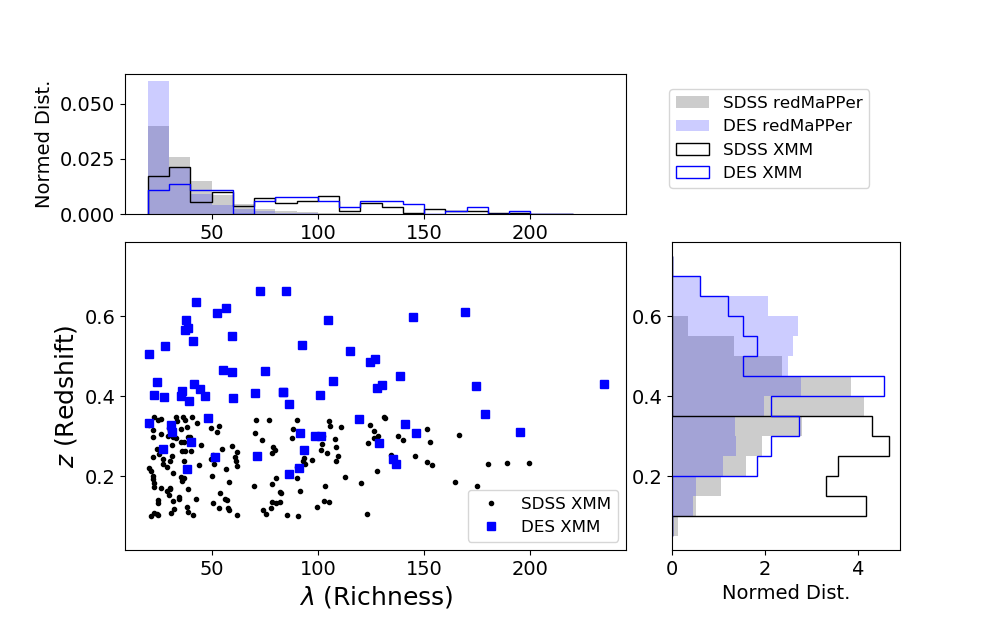}
 \caption{Redshift and richness distributions of the SDSS (black) and DES (blue) redMaPPer clusters matched to archival XMM observations.}
 \label{fig:xray-redmapper}
\end{figure}


We further apply off axis angle and SNR cuts to the XMM samples.  The SNR was determined in the same way as for the {\em Chandra} sample i.e. within $500h^{-1}$ kpc, using 0.5 --2.0 keV {\em XMM} images.  For the XMM-SDSS sample, we require the detections S/N to be $>6.5$, and the \RM~centers to be within 8.5 arcmin of the aim point (or 6.5 arcmin away from the FOV edge assuming a 15 arcmin FOV radius). For the XMM-DES sample, we again require the detections S/N to be $>6.5$, but allow the \RM~centers to be up to 10.5 arcmin of the aim point. The FOV cuts ensure that the corresponding \RM~centers are more than 500 kpc away from the FOV edge at $z=0.1$ for the SDSS sample, or at $z=0.2$ for the DES sample. The S/N cuts were imposed to match the 6.5 S/N cuts in the Chandra analysis.
With these FOV and S/N cuts, the final SDSS and DES XMM samples contain 163 and 66 clusters respectively. Figure~\ref{fig:xray-redmapper} shows their richness and mass distributions.

Further details of the \RM--XMM joint sample development, the peak measurements,  the signal to noise estimation and the individual mis-percolation cases can be found in Giles et al. (in prep). 


\subsection{The XMM-Chandra and XMM-redMaPPer Offsets}
\label{XMMChandra}
\begin{figure}
\includegraphics[width=0.95\linewidth]{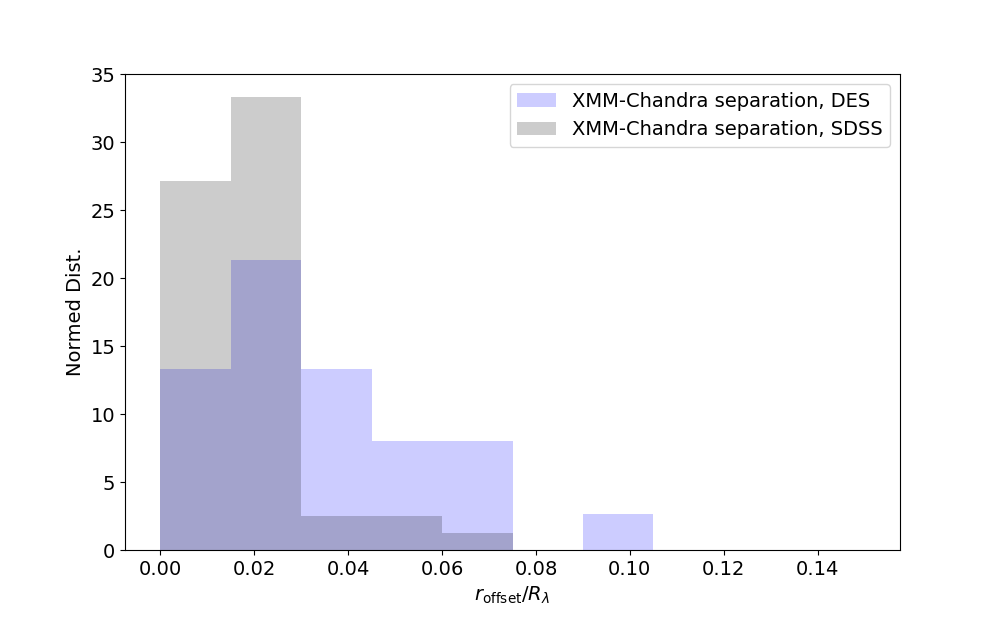}
\caption{The $R_\lambda$ ($R_\lambda =(\lambda/100)^{0.2}  h^{-1}\mpc$) scaled offset distribution between the Chandra and XMM peak identifications for the same redMaPPer clusters.}
\label{fig:xcs_chandra}
\end{figure}

A subsample of the \RM~clusters, 54 in the SDSS sample, and 25 in the DES sample, are analyzed by both the XMM and Chandra analyses. With these overlapping cases, we compare the XMM peak measurements to those from Chandra. Figure~\ref{fig:xcs_chandra} shows the offset distribution between XMM and Chandra peak identifications for the same \RM~clusters, scaled by their $R_\lambda$. The XMM and Chandra peak identifications are highly consistent: their separations are within 0.05 $R_\lambda$  for 53/54 of the overlapping SDSS clusters, and 20/25 of the overlapping DES clusters. The separations have a wider distribution for the DES \RM~sample reflecting its higher redshift range, and hence higher X-ray peak identification uncertainties  in terms of physical distances.

\begin{figure}
\includegraphics[width=0.95\linewidth]{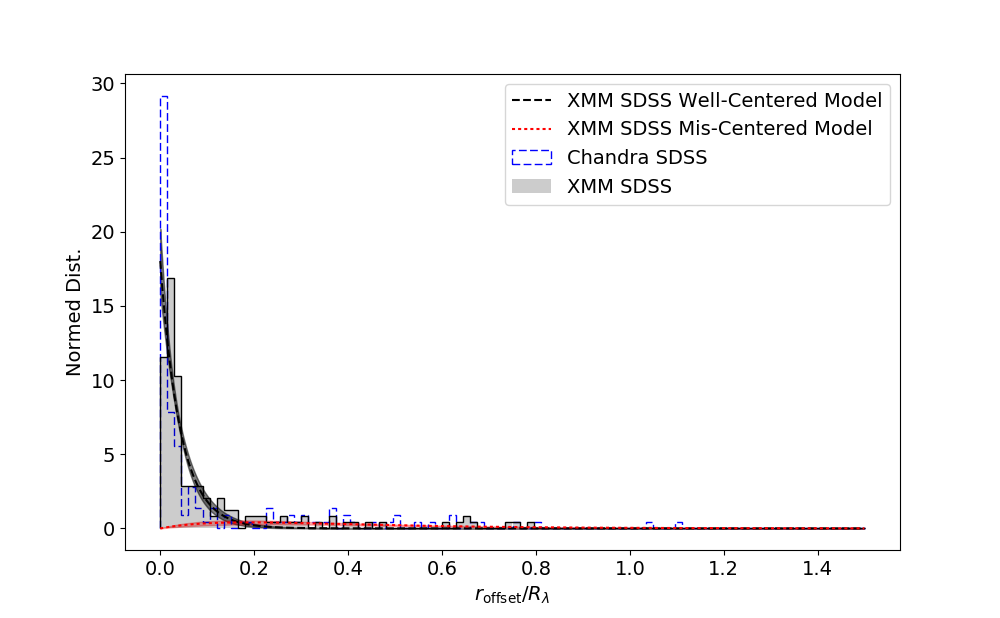}
\caption{The $R_\lambda$ ($R_\lambda =(\lambda/100)^{0.2}  h^{-1}\mpc$) scaled offset distribution between the redMaPPer centers and the X-ray emission peaks for the redMaPPer SDSS samples from the XMM archival observations. The distribution can be fitted with two components -- a concentrated component that represents the well centered redMaPPer clusters, and an extended component that represents the mis-centered redMaPPer clusters. The best fit SDSS offset model is shown as the solid lines (black: well-centered model, red: mis-centered model), with the shaded regions representing the uncertainties. As a comparison, we also show the corresponding offset distribution from the analysis with Chandra archival data (blue dashed histogram).}
\label{fig:xmmSDSS}
\end{figure}

\begin{figure}
\includegraphics[width=0.95\linewidth]{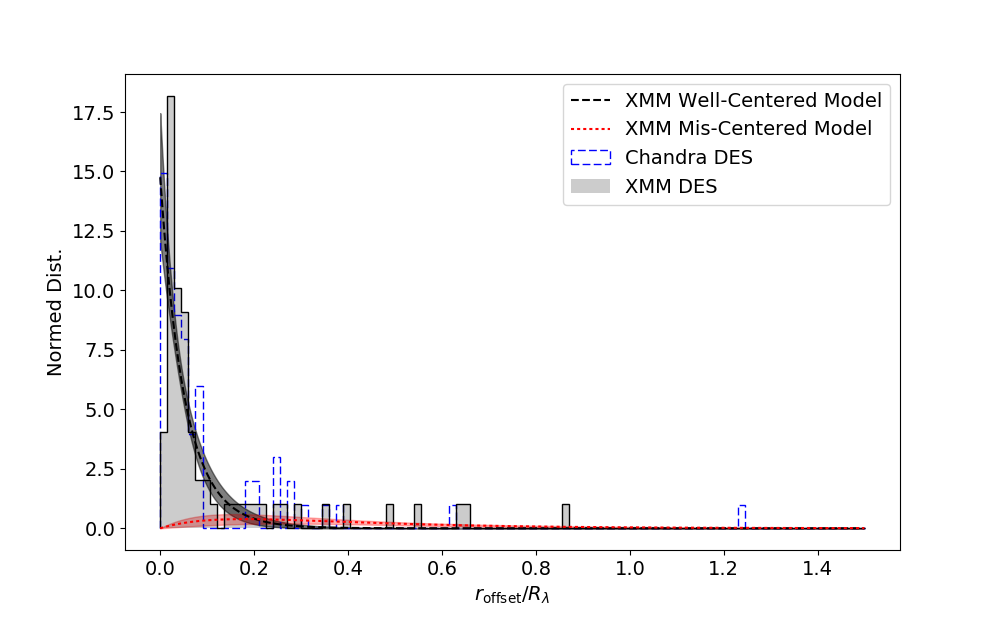}
\caption{The $R_\lambda$ ($R_\lambda =(\lambda/100)^{0.2}  h^{-1}\mpc$) scaled offset distribution between the redMaPPer centers and the X-ray emission peaks for the redMaPPer DES samples from the XMM archival observations. The distribution can be fitted with two components -- a concentrated component that represents the well centered redMaPPer clusters, and an extended component that represents the mis-centered redMaPPer clusters. The best fit DES offset model is shown as the solid lines (black: well-centered model, red: mis-centered model), with the shaded regions representing the uncertainties. As a comparison, we also show the corresponding offset distribution from the analysis with Chandra archival data (blue dashed histogram).}
\label{fig:xmmDES}
\end{figure}

\begin{figure}
\includegraphics[width=0.95\linewidth]{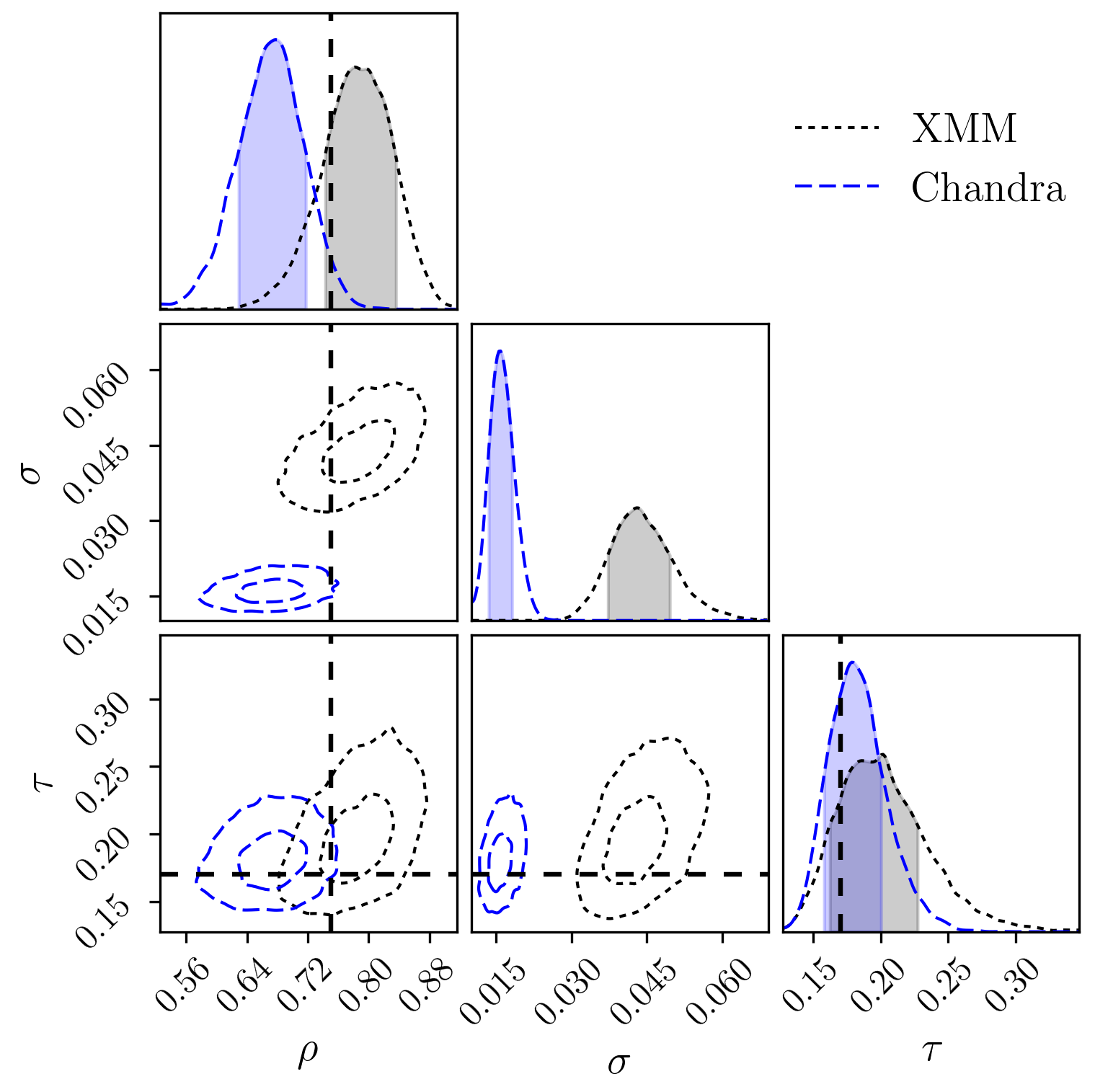}
\caption{Centering offset parameter constraints (Equation~\ref{eq:offset}) for the XMM SDSS (gray) redMaPPer samples. $76\pm6\%$ of the SDSS redMaPPer clusters appear to be well centered (indicated by the $\rho$ parameter). For the mis-centered clusters, their mis-centering offsets is characterized by a Gamma distribution with a characteristic offset (the $\tau$ parameter) of $0.16\pm0.03$ $R_\lambda$.  As a comparison, we also show the corresponding model constraints from the SDSS analysis with Chandra archival data (blue). The prior central values adopted in DES cosmological analyses as described in Section~\ref{sec:cosmo} are shown as the black dashed lines.}
\label{fig:xmmSDSS_mcmc}
\end{figure}

\begin{figure}
\includegraphics[width=0.95\linewidth]{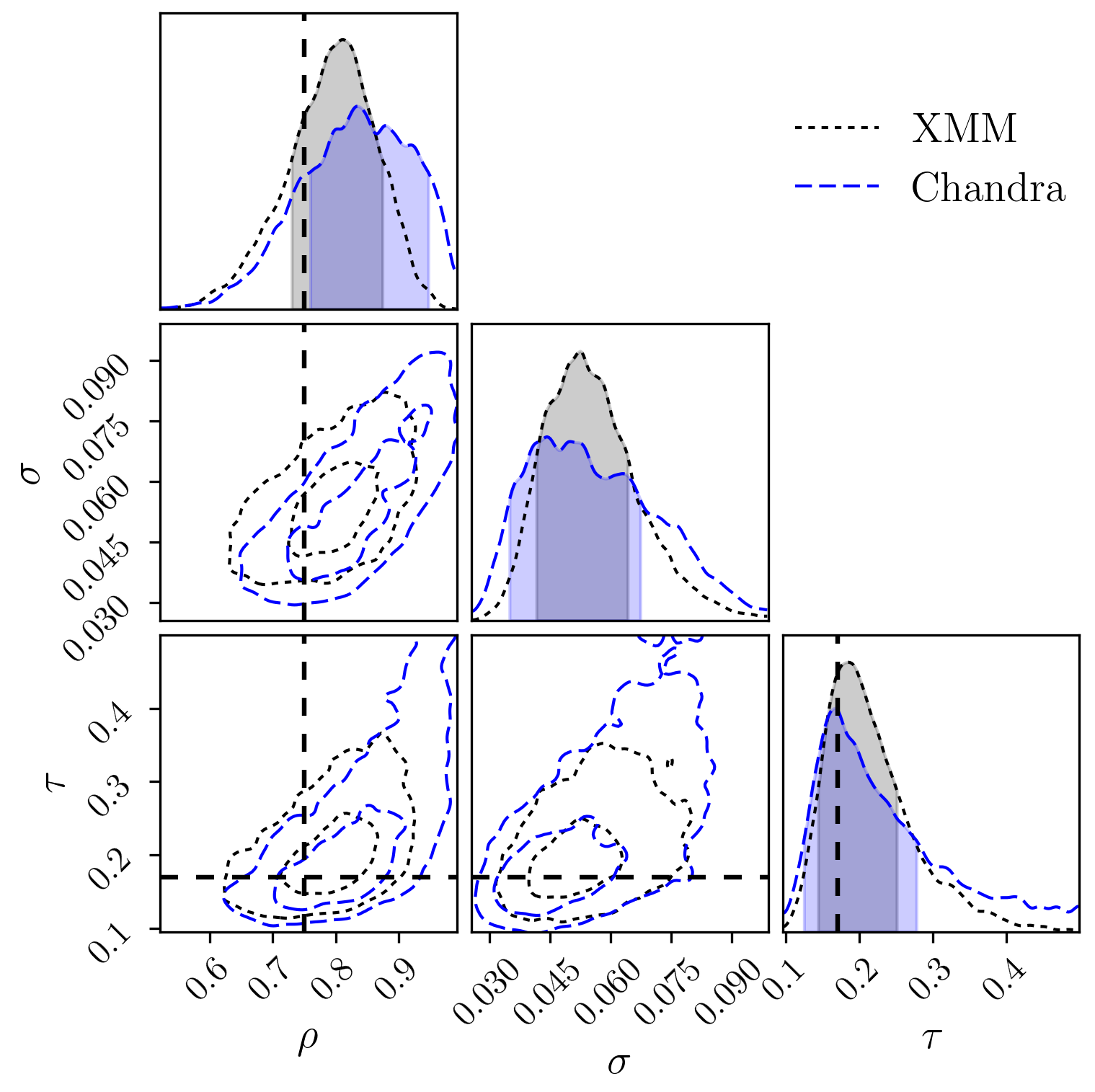}
\caption{Centering offset parameter constraints (Equation~\ref{eq:offset}) for the XMM DES (gray) redMaPPer samples. $74\pm10\%$ of the DES redMaPPer clusters appear to be well centered (indicated by the $\rho$ parameter). For the mis-centered clusters, their mis-centering offsets is characterized by a Gamma distribution with a characteristic offset (the $\tau$ parameter) of $0.20\pm0.07$ $R_\lambda$.  As a comparison, we also show the corresponding model constraints from the DES analysis with Chandra archival data (blue). The prior central values adopted in DES cosmological analyses as described in Section~\ref{sec:cosmo} are shown as the black dashed lines.}
\label{fig:xmmDES_mcmc}
\end{figure}

Given the consistency between XMM and Chandra X-ray peak identifications, we constrain the \RM~centering offset model proposed in Section~\ref{sec:center} with the XMM peaks. Figure~\ref{fig:xmmSDSS} and~\ref{fig:xmmDES} respectively show the offset distributions between the XMM peaks and the \RM~centers for the SDSS and DES samples, with comparisons to their corresponding Chandra offset distributions. Figure~\ref{fig:xmmSDSS_mcmc}, ~\ref{fig:xmmDES_mcmc} and Table~\ref{tbl:centering_params_XMM} show the model parameter constraints from these XMM samples. The $\sigma$ parameter in our \RM~centering offset model (Equation~\ref{eq:offset}) represents the X-ray peak offset to cluster central galaxy for well-centered clusters, which is further smeared by X-ray peak identification uncertainty and X-ray telescope PSFs. Since $\sigma$ is in the unit of physical distance, the $\sigma$ difference can be driven by the different angular resolutions of XMM and Chandra at low redshift, and other x-ray peak identification uncertainties, and x-ray peak-galaxy center separations at higher redshift. Hence, we do not expect this parameter to agree between XMM and Chandra, and the $\sigma$ difference is especially larger for the lower redshift SDSS \RM~samples. For the other two parameters of the model, $\rho$ and $\tau$, which respectively represent the well-centered fractions of \RM~and the centering offset of mis-centered \RM~clusters, the constraints from XMM are consistent with those from Chandra for both the SDSS and DES \RM~samples within 2 standard deviations. 

Notably, no selection cuts in the XMM analysis were made to make its centering model constraints better match the Chandra results, yet their centering offset results are consistent with each other. We conclude that the \RM~mis-centering offset modeling presented in this paper are robust upon investigation with archival XMM data.

\begin{table}
\caption{Centering offset Parameter constraints (Equation~\ref{eq:offset}) for the XMM DES and SDSS redMaPPer samples.}
\label{tbl:centering_params_XMM}
\vspace{1em}
\begin{tabular}{lllll}
\hline
 & $\rho$  & $\sigma$  & $\tau$\\
  \vspace{1em}
Prior & $[0.3, 1]$ & $[0.0001, 0.1]$ & $[0.08, 0.5]$ \\
  \hspace{0.5em} \\
\hline
\\
XMM SDSS Posterior & $0.781\substack{+0.055 \\ -0.038}$ &  $0.0432\substack{+0.0063\\ -0.0059}$  &  $0.201\substack{+0.026 \\ -0.039}$ \\
  \hspace{0.5em} \\
\hline
\\
XMM DES Posterior & $0.815\substack{+0.059 \\ -0.085}$ &  $0.053\substack{+0.012\\ -0.011}$  &  $0.185\substack{+0.066\\ -0.041}$ \\
  \hspace{0.5em} \\
\hline
 \\
\end{tabular}
\end{table}

\section{Summary}
\label{sec:conclude}

This analysis makes use of the archival X-ray observations to constrain the centering performance of the \RM~cluster finding algorithm. We calibrate the well-centered fraction of \RM~clusters for both the SDSS and DES samples with the X-ray emission peaks. The offsets between the \RM~centers and X-ray peaks are well modeled by a two component distribution, which indicates that $69^{+3.5}_{-5.1}\%$ and $83.5^{+11.2}_{-7.5}\%$ of the clusters are well centered in the SDSS and DES samples. The offset distribution of the mis-centered \RM~clusters are modeled with a Gamma distribution, and cluster mass modeling appears to be more sensitive to the accuracy of these mis-centering offsets than the mis-centering fraction. 

With the upcoming DES Year 3 and Year 5 data, we expect the \RM~centering constraints to continue improving with $\sim$ 2 times larger overlapping  samples between DES and archival X-ray observations, which may permit us to quantify the dependence of the centering parameters on cluster properties, such as cluster richness, redshift, X-ray temperature and luminosity. The current improvement has already lowered the cluster weak lensing mass modeling uncertainties due to mis-centering, to the extent of being in-substantial comparing to the other modeling systematic effects. Since mis-centering is often assumed to be uncorrelated with cluster mass distributions in weak lensing analyses, with the anticipated level of improvement, one may wish to  investigate the correlations of mis-centering with other cluster lensing systematic effects, such as cluster mass modeling uncertainties, cluster orientatoin, triaxiality and projection \citep{McClintock2018}.

The cluster richness estimates tend to be biased lower by mis-centering. In this paper, we propose a richness bias model to describe the effect, which is validated by X-ray centered richness measurements. The richness bias is offset dependent, low for clusters with small mis-centering offset, but larger than 50\% for severely mis-centered clusters. Cluster cosmology studies based on full depth DES data or LSST data should explicitly account for this effect to avoid biased cosmological parameter inferences. 

Code used in this analysis is available from \href{https://github.com/yyzhang/center_modeling_y1}{https://github.com/yyzhang/center\_modeling\_y1}.

\section*{Acknowledgements}
\label{sec:acknowledgements}
TJ is pleased to acknowledge funding support from DE-SC0010107 and DE-SC0013541. KR, PG and SB acknowledge support from the UK Science and Technology Facilities Council via grants ST/P000252/1 and ST/N504452/1, respectively. DG was supported in part by the U.S. Department of Energy under contract number DE-AC02-76SF00515 and by Chandra Award Number GO8-19101A, issued by the Chandra X-ray Observatory Center. We use Monte Carlo Markov Chain sampling in this analyses, which are performed with the PYMC \citep{Patil10pymc:bayesian} and emcee \citep{2013PASP..125..306F} python packages.

Funding for the DES Projects has been provided by the U.S. Department of Energy, the U.S. National Science Foundation, the Ministry of Science and Education of Spain, 
the Science and Technology Facilities Council of the United Kingdom, the Higher Education Funding Council for England, the National Center for Supercomputing 
Applications at the University of Illinois at Urbana-Champaign, the Kavli Institute of Cosmological Physics at the University of Chicago, 
the Center for Cosmology and Astro-Particle Physics at the Ohio State University,
the Mitchell Institute for Fundamental Physics and Astronomy at Texas A\&M University, Financiadora de Estudos e Projetos, 
Funda{\c c}{\~a}o Carlos Chagas Filho de Amparo {\`a} Pesquisa do Estado do Rio de Janeiro, Conselho Nacional de Desenvolvimento Cient{\'i}fico e Tecnol{\'o}gico and 
the Minist{\'e}rio da Ci{\^e}ncia, Tecnologia e Inova{\c c}{\~a}o, the Deutsche Forschungsgemeinschaft and the Collaborating Institutions in the Dark Energy Survey. 

The Collaborating Institutions are Argonne National Laboratory, the University of California at Santa Cruz, the University of Cambridge, Centro de Investigaciones Energ{\'e}ticas, 
Medioambientales y Tecnol{\'o}gicas-Madrid, the University of Chicago, University College London, the DES-Brazil Consortium, the University of Edinburgh, 
the Eidgen{\"o}ssische Technische Hochschule (ETH) Z{\"u}rich, 
Fermi National Accelerator Laboratory, the University of Illinois at Urbana-Champaign, the Institut de Ci{\`e}ncies de l'Espai (IEEC/CSIC), 
the Institut de F{\'i}sica d'Altes Energies, Lawrence Berkeley National Laboratory, the Ludwig-Maximilians Universit{\"a}t M{\"u}nchen and the associated Excellence Cluster Universe, 
the University of Michigan, the National Optical Astronomy Observatory, the University of Nottingham, The Ohio State University, the University of Pennsylvania, the University of Portsmouth, 
SLAC National Accelerator Laboratory, Stanford University, the University of Sussex, Texas A\&M University, and the OzDES Membership Consortium.

Based in part on observations at Cerro Tololo Inter-American Observatory, National Optical Astronomy Observatory, which is operated by the Association of 
Universities for Research in Astronomy (AURA) under a cooperative agreement with the National Science Foundation.

The DES data management system is supported by the National Science Foundation under Grant Numbers AST-1138766 and AST-1536171.
The DES participants from Spanish institutions are partially supported by MINECO under grants AYA2015-71825, ESP2015-66861, FPA2015-68048, SEV-2016-0588, SEV-2016-0597, and MDM-2015-0509, 
some of which include ERDF funds from the European Union. IFAE is partially funded by the CERCA program of the Generalitat de Catalunya.
Research leading to these results has received funding from the European Research
Council under the European Union's Seventh Framework Program (FP7/2007-2013) including ERC grant agreements 240672, 291329, and 306478.
We  acknowledge support from the Australian Research Council Centre of Excellence for All-sky Astrophysics (CAASTRO), through project number CE110001020.

This manuscript has been authored by Fermi Research Alliance, LLC under Contract No. DE-AC02-07CH11359 with the U.S. Department of Energy, Office of Science, Office of High Energy Physics. The United States Government retains and the publisher, by accepting the article for publication, acknowledges that the United States Government retains a non-exclusive, paid-up, irrevocable, world-wide license to publish or reproduce the published form of this manuscript, or allow others to do so, for United States Government purposes.

\section*{Affiliations}
{\small
$^{1}$ Fermi National Accelerator Laboratory, P. O. Box 500, Batavia, IL 60510, USA\\
$^{2}$ Santa Cruz Institute for Particle Physics, Santa Cruz, CA 95064, USA\\
$^{3}$ Department of Physics, University of Arizona, Tucson, AZ 85721, USA\\
$^{4}$ Department of Physics, Carnegie Mellon University, Pittsburgh, Pennsylvania 15312, USA\\
$^{5}$ Department of Physics, University of Michigan, Ann Arbor, MI 48109, USA\\
$^{6}$ Department of Physics and Astronomy, Pevensey Building, University of Sussex, Brighton, BN1 9QH, UK\\
$^{7}$ Kavli Institute for Particle Astrophysics \& Cosmology, P. O. Box 2450, Stanford University, Stanford, CA 94305, USA\\
$^{8}$ SLAC National Accelerator Laboratory, Menlo Park, CA 94025, USA\\
$^{9}$ Department of Astronomy, University of Michigan, Ann Arbor, MI 48109, USA\\
$^{10}$ Faculty of Physics, Ludwig-Maximilians-Universit\"at, Scheinerstr. 1, 81679 Munich, Germany\\
$^{11}$ Department of Physics, Stanford University, 382 Via Pueblo Mall, Stanford, CA 94305, USA\\
$^{12}$ Department of Physics and Astronomy, Stony Brook University, Stony Brook, NY 11794, USA\\
$^{13}$ Universit\"ats-Sternwarte, Fakult\"at f\"ur Physik, Ludwig-Maximilians Universit\"at M\"unchen, Scheinerstr. 1, 81679 M\"unchen, Germany\\
$^{14}$ Astrophysics Research Institute, Liverpool John Moores University, IC2, Liverpool Science Park, 146 Brownlow Hill, Liverpool, L3 5RF, UK\\
$^{15}$ Astrophysics \& Cosmology Research Unit, School of Mathematics, Statistics \& Computer Science, University of KwaZulu-Natal, Westville Campus, Durban 4041, South Africa\\
$^{16}$ Max Planck Institute for Extraterrestrial Physics, Giessenbachstrasse, 85748 Garching, Germany\\
$^{17}$ Institute for Astronomy, University of Edinburgh, Royal Observatory, Blackford Hill, Edinburgh EH9 3NJ\\
$^{18}$ Institute for Astronomy, University of Edinburgh, Edinburgh EH9 3HJ, UK\\
$^{19}$Department of Physics and Astronomy, Uppsala University, Box 516, SE-751 20  Uppsala, Sweden\\
$^{20}$ Sub-department of Astrophysics, Department of Physics, University of Oxford, Denys Wilkinson Building, Keble Road, Oxford OX1 3RH, UK and Department of Physics, Lancaster University, Lancaster LA1 4 YB, UK\\
$^{21}$ Instituto de Astrofisica e Ciencias do Espaco, Universidade do Porto, CAUP, Rua das Estrelas, P-4150-762 Porto, Portugal and Departamento de Fisica e Astronomia, Faculdade de Ciencias, Universidade do Porto, Rua do Campo Alegre, 687, P-4169-007 Porto, Portugal\\
$^{22}$ Institute of Cosmology and Gravitation, University of Portsmouth, Portsmouth, PO1 3FX, UK\\
$^{23}$ LSST, 933 North Cherry Avenue, Tucson, AZ 85721, USA\\
$^{24}$ Physics Department, 2320 Chamberlin Hall, University of Wisconsin-Madison, 1150 University Avenue Madison, WI  53706-1390\\
$^{25}$ CNRS, UMR 7095, Institut d'Astrophysique de Paris, F-75014, Paris, France\\
$^{26}$ Sorbonne Universit\'es, UPMC Univ Paris 06, UMR 7095, Institut d'Astrophysique de Paris, F-75014, Paris, France\\
$^{27}$ Department of Physics \& Astronomy, University College London, Gower Street, London, WC1E 6BT, UK\\
$^{28}$ Centro de Investigaciones Energ\'eticas, Medioambientales y Tecnol\'ogicas (CIEMAT), Madrid, Spain\\
$^{29}$ Laborat\'orio Interinstitucional de e-Astronomia - LIneA, Rua Gal. Jos\'e Cristino 77, Rio de Janeiro, RJ - 20921-400, Brazil\\
$^{30}$ Department of Astronomy, University of Illinois at Urbana-Champaign, 1002 W. Green Street, Urbana, IL 61801, USA\\
$^{31}$ National Center for Supercomputing Applications, 1205 West Clark St., Urbana, IL 61801, USA\\
$^{32}$ Institut de F\'{\i}sica d'Altes Energies (IFAE), The Barcelona Institute of Science and Technology, Campus UAB, 08193 Bellaterra (Barcelona) Spain\\
$^{33}$ Institut d'Estudis Espacials de Catalunya (IEEC), 08034 Barcelona, Spain\\
$^{34}$ Institute of Space Sciences (ICE, CSIC),  Campus UAB, Carrer de Can Magrans, s/n,  08193 Barcelona, Spain\\
$^{35}$ Observat\'orio Nacional, Rua Gal. Jos\'e Cristino 77, Rio de Janeiro, RJ - 20921-400, Brazil\\
$^{36}$ Department of Physics, IIT Hyderabad, Kandi, Telangana 502285, India\\
$^{37}$ Excellence Cluster Universe, Boltzmannstr.\ 2, 85748 Garching, Germany\\
$^{38}$ Kavli Institute for Cosmological Physics, University of Chicago, Chicago, IL 60637, USA\\
$^{39}$ Instituto de Fisica Teorica UAM/CSIC, Universidad Autonoma de Madrid, 28049 Madrid, Spain\\
$^{40}$ Department of Physics, ETH Zurich, Wolfgang-Pauli-Strasse 16, CH-8093 Zurich, Switzerland\\
$^{41}$ Center for Cosmology and Astro-Particle Physics, The Ohio State University, Columbus, OH 43210, USA\\
$^{42}$ Department of Physics, The Ohio State University, Columbus, OH 43210, USA\\
$^{43}$ Department of Astronomy/Steward Observatory, 933 North Cherry Avenue, Tucson, AZ 85721-0065, USA\\
$^{44}$ Australian Astronomical Optics, Macquarie University, North Ryde, NSW 2113, Australia\\
$^{45}$ Departamento de F\'isica Matem\'atica, Instituto de F\'isica, Universidade de S\~ao Paulo, CP 66318, S\~ao Paulo, SP, 05314-970, Brazil\\
$^{46}$ George P. and Cynthia Woods Mitchell Institute for Fundamental Physics and Astronomy, and Department of Physics and Astronomy, Texas A\&M University, College Station, TX 77843,  USA\\
$^{47}$ Department of Astrophysical Sciences, Princeton University, Peyton Hall, Princeton, NJ 08544, USA\\
$^{48}$ Instituci\'o Catalana de Recerca i Estudis Avan\c{c}ats, E-08010 Barcelona, Spain\\
$^{49}$ School of Physics and Astronomy, University of Southampton,  Southampton, SO17 1BJ, UK\\
$^{50}$ Brandeis University, Physics Department, 415 South Street, Waltham MA 02453\\
$^{51}$ Computer Science and Mathematics Division, Oak Ridge National Laboratory, Oak Ridge, TN 37831\\
$^{52}$ Argonne National Laboratory, 9700 South Cass Avenue, Lemont, IL 60439, USA\\
}

\bibliographystyle{mnras}
\bibliography{references}

\bsp	
\label{lastpage}
\end{document}

%% file: centering_authors.tex
\author[DES Collaboration]{
\parbox{\textwidth}{
\Large
Y.~Zhang$^{1}$,
T.~Jeltema$^{2}$,
D.~L.~Hollowood$^{2}$,
S.~Everett$^{2}$,
E.~Rozo$^{3}$,
A.~Farahi$^{4,5}$,
A.~Bermeo$^{6}$,
S.~Bhargava$^{6}$,
P.~Giles$^{6}$,
A.~K.~Romer$^{6}$,
R.~Wilkinson$^{6}$,
E.~S.~Rykoff$^{7,8}$,
A.~Mantz$^{7}$,
H.~T.~Diehl$^{1}$,
A.~E.~Evrard$^{9,5}$,
C.~Stern$^{10}$,
D.~Gruen$^{11,7,8}$,
A.~von der Linden$^{12}$,
M.~Splettstoesser$^{6}$,
X.~Chen$^{5}$,
M.~Costanzi$^{13}$,
S.~Allen$^{11}$,
C.~Collins$^{14}$,
M.~Hilton$^{15}$,
M.~Klein$^{10,16}$,
R.G.~Mann$^{17}$,
M.~Manolopoulou$^{18}$,
G.~Morris$^{8}$,
J.~Mayers$^{6}$,
M.~Sahlen$^{19}$,
J.~Stott$^{20}$,
C.~Vergara Cervantes$^{6}$,
P.T.P.~Viana$^{21}$,
R.~H.~Wechsler$^{8}$,
S.~Allam$^{1}$,
S.~Avila$^{22}$,
K.~Bechtol$^{23,24}$,
E.~Bertin$^{25,26}$,
D.~Brooks$^{27}$,
D.~L.~Burke$^{7,8}$,
A.~Carnero~Rosell$^{28,29}$,
M.~Carrasco~Kind$^{30,31}$,
J.~Carretero$^{32}$,
F.~J.~Castander$^{33,34}$,
L.~N.~da Costa$^{29,35}$,
J.~De~Vicente$^{28}$,
S.~Desai$^{36}$,
J.~P.~Dietrich$^{37,10}$,
P.~Doel$^{27}$,
B.~Flaugher$^{1}$,
P.~Fosalba$^{33,34}$,
J.~Frieman$^{1,38}$,
J.~Garc\'ia-Bellido$^{39}$,
E.~Gaztanaga$^{33,34}$,
R.~A.~Gruendl$^{30,31}$,
J.~Gschwend$^{29,35}$,
G.~Gutierrez$^{1}$,
W.~G.~Hartley$^{27,40}$,
K.~Honscheid$^{41,42}$,
B.~Hoyle$^{16,13}$,
E.~Krause$^{43}$,
K.~Kuehn$^{44}$,
N.~Kuropatkin$^{1}$,
M.~Lima$^{45,29}$,
M.~A.~G.~Maia$^{29,35}$,
J.~L.~Marshall$^{46}$,
P.~Melchior$^{47}$,
F.~Menanteau$^{30,31}$,
C.~J.~Miller$^{9,5}$,
R.~Miquel$^{48,32}$,
R.~L.~C.~Ogando$^{29,35}$,
A.~A.~Plazas$^{47}$,
E.~Sanchez$^{28}$,
V.~Scarpine$^{1}$,
R.~Schindler$^{8}$,
S.~Serrano$^{33,34}$,
I.~Sevilla-Noarbe$^{28}$,
M.~Smith$^{49}$,
M.~Soares-Santos$^{50}$,
E.~Suchyta$^{51}$,
M.~E.~C.~Swanson$^{31}$,
G.~Tarle$^{5}$,
D.~Thomas$^{22}$,
D.~L.~Tucker$^{1}$,
V.~Vikram$^{52}$,
W.~Wester$^{1}$
\begin{center} (DES Collaboration) \end{center}
\bigskip
\small{{\em Author affiliations are listed at the end of this paper.}}
}
}